\documentclass[manuscript]{aastex63}

\received{}
\revised{}
\accepted{}
\submitjournal{ICARUS}

\shorttitle{Dynamical origin of the Dwarf Planet Ceres}
\shortauthors{Ribeiro et al.}

\begin{document}

\title{Dynamical origin of the Dwarf Planet Ceres  \footnote{Article accepted for publication on  }}

 \correspondingauthor{Rafael Ribeiro de Sousa}
 \email{rafanw72@gmail.com, rribeiro@oac.eu}

\author{Rafael Ribeiro de Sousa}
\affiliation{S\~ao Paulo State University, UNESP,Campus of Guaratinguet\'a, Av. Dr. Ariberto Pereira da Cunha, 333 - Pedregulho, Guaratinguet\'a - SP, 12516-410, Brazil}
%

%
%

%
 \author{Alessandro Morbidelli}
 \affil{ Laboratoire Lagrange, UMR7293, Universit\'e C\^ote d'Azur, CNRS, Observatoire de la C\^ote d'Azur, Boulevard de l'Observatoire, 06304 Nice Cedex 4, France}
 
  \author{Rodney Gomes}
 \affil{Observat\'orio Nacional, Rua General Jos\'e Cristino 77, CEP 20921-400, Rio de Janeiro, RJ,  Brazil}
 \email{rodney@on.br}

\author{Ernesto Vieira Neto}
\affiliation{S\~ao Paulo State University, UNESP,Campus of Guaratinguet\'a, Av. Dr. Ariberto Pereira da Cunha, 333 - Pedregulho, Guaratinguet\'a - SP, 12516-410, Brazil}

 \author{Andre Izidoro}
 \affiliation{Department of Physics and Astronomy  6100 MS 550, Rice University, Houston, TX 77005, USA}
 \affiliation{Department of Earth, Environmental and Planetary Sciences, 6100 MS 126, Rice University, Houston, TX 77005, USA}
 \affiliation{S\~ao Paulo State University, UNESP,Campus of Guaratinguet\'a, Av. Dr. Ariberto Pereira da Cunha, 333 - Pedregulho, Guaratinguet\'a - SP, 12516-410, Brazil}

\author{Abreu\c{c}on Atanasio Alves}
\affiliation{S\~ao Paulo State University, UNESP,Campus of Guaratinguet\'a, Av. Dr. Ariberto Pereira da Cunha, 333 - Pedregulho, Guaratinguet\'a - SP, 12516-410, Brazil}




\begin{abstract}
The Dwarf Planet Ceres revealed the presence of ammonia and other unique properties compared to other asteroids in the main belt which suggests that it was not formed \textit{in situ}.
We model the early dynamical evolution of the outer Solar System to study possible dynamical mechanisms to implant a Ceres-sized planetesimal in the asteroid belt from the trans-Saturnian region. 
We calculate that the fraction of the population of Ceres-sized planetesimals that are captured in the asteroid belt is in the range of $2.8\times10^{-5}$  to $1.2\times10^{-3}$ depending on the initial location
in the outer planetesimal disk. The captured bodies have a 70\% probability to have a semimajor axis between $2.5$ and $3$~au, a 33\% probability to have an eccentricity smaller 
than 0.2 and a 45\% probability to have an orbital inclination smaller than 10$^{\circ}$. Assuming the existence of 3,600 Ceres-size planetesimals in 
the inner part of the trans-Saturnian disk, consistent with the estimate of \citet{Nesvorny2016a} for the trans-Neptunian disk, our estimated capture probability 
and a final 80\% depletion of the asteroid belt during the subsequent giant planet instability, lead to capture $\sim$ 1 Ceres in the asteroid belt, with a probability of 15\%, 34\%, and 51\% to be located 
in the inner, middle and outer belt respectively.
\end{abstract}



\section{Introduction} \label{sec:intro}

The small-body populations in the Solar System provide constraints to unveil the events that happened in its early history.
The asteroids' main belt is the nearest and the best characterized population that we can use to test the dynamical models for the evolution and formation of the planets.
Ceres is the largest object of the main belt (mean physical radius of 470 km) with many interesting physical and chemical properties. Ceres comprises about $0.33\%$ of the total mass
of the asteroid belt and has low density (2.2~g/cm$^{3}$), low albedo and a spectral reflectance similar to carbonaceous chondrites \citep{MccordSotin2005}.
Ceres orbits the Sun deep in the main belt, with a semimajor  axis of $2.767$~au, an eccentricity of $0.097$ and an inclination of about 9.73$^{\circ}$.
The position of Ceres in the main belt is close to many C-types asteroids, but Ceres is not a C-type \citep{1988Bell,2008DPSM}. Ceres is similar 
to G-type asteroids, which are all hydrated whereas C-types show hydration features only in 60\% of the cases \citep{1992IcarB} and is 
rich in ammonia compounds that are typically not found in C-type asteroids and carbonaceous chondrites \citep{2015D}.

Ceres is in rotational hydrostatic equilibrium, at least partially differentiated \citep{2020Neumann} but it did not dynamically sculpt the asteroid population 
in its neighborhood; thus, it is classified as a dwarf planet (although it was considered a planet when it was discovered).
The interior structure of Ceres comprises a water-ice-rich mantle over a rock core, with a bulk rock-ice fraction of 75\%-25\% respectively \citep{2005T}.
All these properties, deduced from the observations of the Dawn mission suggest that Ceres and the Kuiper belt share the same source location during their accretion phase \citep{2020Neumann}. Indeed, the physical and chemical properties of Ceres share many similarities with
the largest Kuiper Belt Objects (Triton, Eris and the Pluto-Charon binary system).
In particular, the discovery of ammonia on Ceres' surface is a strong chemical evidence that Ceres formed in the outer region of the Solar System \citep{2015D}.
The presence of ammonia could mean that Ceres was formed far beyond the water-snowline where ammonia is readily available. It is unlikely that ammonia-rich material drifted 
from the outer Solar System and was accreted by Ceres \textit{in situ} \citep{2016AJV}.

The so-called \textit{Nice model} invokes a dynamical instability among the giant planets that was triggered by their interactions with the primordial outer
planetesimal disk after the removal of gas from the system \citep{Tsiganisetal2005,Levison2011,NesvornyMorby2012}. \citet{2008DPSM,2012DPSM} proposed that Ceres is a planetesimal 
captured from the primordial trans-Neptunian disk during the events described in the \textit{Nice model}. More precisely, Ceres was dynamically scattered inwards during the 
reorganization of the outer Solar System during the instability phase, then it was implanted in a more massive primordial asteroid belt, and finally its orbit was circularized due to dynamical friction. 
The dynamical justification of this proposition comes from models showing KBOs and comets implanted in large numbers into the Jovian Trojan clouds and in the outer asteroid belt \citep{Morbidellietal2005,2009NaturL}.

\citet{2016AJV} studied the capture of trans-Neptunian planetesimals into the asteroid belt using a modified version of the Nice model \citep{NesvornyMorby2012}, which assumes 
that an additional giant planet originally existed in the outer solar system (Jupiter, Saturn, Uranus, Neptune and a Neptune-like body) and also a primordial
disk containing thousands of Pluto-mass bodies \citep{Nesvorny2016a,2021Icar..35613793A}. They were one of the first to investigate how the interactions between the fifth giant planet, 
Jupiter and  planetesimals disk helped to implant planetesimals into the asteroid belt, via temporary captures into the first-order mean motion resonances with Jupiter, while others have 
repeated this study with greater complexity \citep{Quarles_Kaib_2019, Clement2019b, RIBEIRODESOUSA201989}.
Overall, the distribution of trans-Neptunian planetesimals captured in the asteroid belt in the \citet{2016AJV} model reproduces 
the observed distribution of D and P-type asteroids better than in previous models.

\citet{2016AJV} also estimated the probability of Ceres being captured from the trans-Neptunian planetesimal disk (between 23 and 30 au) in the Nice model. 
Assuming that the trans-Neptunian disk once contained 3,600-14,000 Ceres-sized objects \citep{Nesvorny2016a, 2021Icar..35613793A}, \citet{2016AJV} concluded that the probability 
that one was captured into the asteroid belt is too low to be a viable explanation for the origin of Ceres. 

\citet{raymond17} also studied planetesimal capture into the asteroid belt during the accretion phase of the giant planets, when gas was still present. As a consequence of the combined effects of orbital destabilization, gravitational 
scattering and aerodynamic gas drag, planetesimals from the Jupiter-Saturn region could be implanted into the asteroid belt. These implanted asteroids preferentially land into the outer main belt, 
and their source region was 5~au-wide (from 4 to 9~au). Accounting for the migration of Jupiter and Saturn from initially larger distances broadens the source region of implanted objects up to 10 -- 20 au. \citet{raymond17} 
identified the implanted planetesimals with the C-type asteroids of the main belt, because cosmochemical evidence \citep{Kruijer2017} shows that these bodies should have formed beyond Jupiter. However, other asteroid types 
(e.g. G, D, P) could in principle also be implanted. \citet{raymond17} also observed that the distribution of $D=1\,000$~km implanted planetesimals peaks at around $2.7$~au. Thus, they claimed that Ceres was implanted during 
the growth of Jupiter and Saturn. However, the amount of original planetesimals in Jupiter-Saturn region is highly uncertain \citep{2020RaymondNesvorny}.

The goal of this paper is to investigate the possibility of capture of Ceres from the Trans-Saturnian region, which has the advantage of drafting Ceres from a farther heliocentric distance, 
where bodies are more likely to be ammonia-rich. In contrast to \citet{raymond17}, we study the capture of Ceres-sized planetesimals after the full 
formation of Jupiter and Saturn but during the growth of Uranus and Neptune. To do this, we use the model of \citet{Izidoroetal2015}, described in Section 2, which is the most successful one to reproduce 
the formation of these two planets and their large obliquities. 
To study the possibility that Ceres is a refugee from the primordial Trans-Saturnian disk we proceed as follows. First, we reproduce the dynamical evolution of the planetary bodies simulated by 
\citet{Izidoroetal2015} and leading to the successful formation of Uranus and Neptune. 
Then, we generate a planetesimal disk containing only objects of Ceres-sized planetesimals. We calculate the fraction of Ceres-sized planetesimals that are placed in orbits with a semimajor axis smaller than 3.5~au and study the mechanism to implant these objects in a permanent stable orbit very close to the current orbit of Ceres.

 \section{Methods}
 \label{Methods}

 \citet{Izidoroetal2015} proposed that the Uranus and Neptune formed in two steps. By the time that Jupiter and Saturn had undergone 
rapid gas accretion to become giant planets, pebble accretion had produced a system of protoplanets of $\sim5$ Earth masses, comparable in mass to the most-frequently observed super-Earths 
\citep[e.g.,][]{mayor11,batalha13,marcy14,wolfgang16}. These protoplanets migrated inward in Type-I regime due to tidal torques from the gas disk \citep{GoldreichTremaine80,Ward86,Tanakaetal2002}.
While it is unclear whether Jupiter and Saturn at this point would have been migrating inward, outward or had roughly stationary orbits 
\citep{MassetSnellgrove2001,MorbidelliCrida2007,PierensNelson2008,PierensRaymond2011,RaymondMorbidelli2014,Pierensetal2014}, the gas giants' migration was certainly slower than that of the protoplanets for typical disk viscosities. 
The protoplanets could not migrate through Jupiter and Saturn's orbits; rather, the gas giants acted as a migration barrier \citep{Izidoroetal2015b}. Protoplanets became trapped in mean motion resonances with Saturn, 
forming a \textit{resonant chain}. As more protoplanets migrated inward, the resonant chain was destabilized, leading to giant (obliquity-generating) collisions and to another phase of migration into a new \textit{resonant chain}. 
\cite{Izidoroetal2015} showed that this process typically produces 2-3 ice giants with masses comparable to those of Uranus and Neptune in resonant chains that include Jupiter and Saturn.

In our previous work \citep{RIBEIRO2020113605}, we concentrated our efforts using one of the best simulations from \citet{Izidoroetal2015} to study the dynamical sculpting of the primordial outer planetesimal disk 
during the accretion of Uranus and Neptune from migrating protoplanets during the gas disk phase. The \citet{RIBEIRO2020113605} model implemented the interpolation of the orbital elements of all massive bodies 
(Jupiter, Saturn and the protoplanets) into the \citet{Izidoroetal2015} simulation using spline functions. The interpolated evolution of the massive bodies was used to simulate the planetesimal evolution 
including the effects of a gaseous protoplanetary disk on the planetesimals (gas drag and dynamical friction; Section~\ref{gas}).

In this work, we use the same approach as in \citet{RIBEIRO2020113605} to model the capture process of Ceres because: 
i) \citet{Izidoroetal2015} scenario is the most successful model in explaining quantitatively the origin of Uranus and Neptune and ii) The method of interpolation 
presented in \citet{RIBEIRO2020113605} has the advantage of implementing the exact same evolution of the giant planets and protoplanets of \citet{Izidoroetal2015} in all simulations (e.g varying the initial location 
of planetesimals).  
Most importantly, the \citet{Izidoroetal2015} simulation showed that some protoplanets temporarily reached the inner region of 
the Solar System. The history of these protoplanets may dynamically affect the capture process of a Ceres-sized planetesimal.
Note that the interpolation model for the evolution of the protoplanets already accounts for the tidal forces between the protoplanets and 
the gas disk \citep{Izidoroetal2015,RIBEIRO2020113605}.

\subsection{Interactions of planetesimals with the gaseous protoplanetary disk}
\label{gas}

As in \citet{RIBEIRO2020113605} we consider two different effects of the gas drag on planetesimals: the aerodynamic gas drag and the gas dynamical friction. 

The aerodynamic gas drag force on a particle moving in a gas disk environment is expressed as a function of its shape, size, velocity and gas conditions. In the particular case of a spherical body with radius $R$, 
the drag force is in a direction opposite to the particle velocity and is expressed by:
\begin{equation}
\vec{F}_D= - \frac{1}{2} C_D \pi R^2 \rho_g v_{rel} \vec{v}_{rel} ,
\end{equation}
where $C_D$ is the drag coefficient and $\vec{v}_{rel}$ is the relative velocity vector between the gas and the planetesimal.
The drag coefficient for a spherical object is a function of the Reynolds number ($R_e$), which is a measure of the turbulence
of the gas in the wake of a planetesimal, the Mach Number ($M$), and Knudsen number ($K$).
To evaluate the drag coefficient we used the same approach of \citet{Brasseretal2007} where they estimated
the values of $M$, $K$ and $R_e$ as a function of planetesimal's size and velocity.

The gaseous dynamical friction (GDF) effect arises from the fact that a perturber moving on a straight line in a uniform gaseous medium generates a wake, which in 
turn exerts a gravitational torque on the perturber. The aerodynamic drag force is dominant for small planetesimals and scales as the square of the radius of the planetesimals. The GDF is 
dominant for intermediate sizes planetesimals and scales with the sixth power of the planetesimal radius. 
\citet{Grishin2015} showed that for planetesimals in the mass range of $m\sim {10}^{21}-{10}^{25}$~g the GDF is possibly dominant over gas-drag 
for some gas density conditions. The GDF force is given by:

\begin{equation}
\vec{F}_{GDF}= - \frac{4 \pi G^2m^2 \rho_g}{v_{rel}^3} \vec{v}_{rel} I(M) ,
\end{equation}
$I(M)$ is a dimensionless factor depending on the Mach number ($M$) and it is given by:

\begin{equation} I(M) =  \\\frac{1}{2}\ln\left(\frac{1+M}{1-M}\right) - M, \quad for\:M < 1 
\end{equation}

\begin{equation} I(M) = \\\frac{1}{2}\ln\left(1-\frac{1}{M^2}\right) + \ln\left(\frac{v_{rel}\:t}{R}\right), \quad for\:M>1
\end{equation}

\noindent where the GDF may be important to investigate the capture of Ceres ($m$ $\sim {10}^{24}$~g). 
For the disk gas properties, we used those adopted in \citet{Izidoroetal2015}, which had been calculated from a hydrodynamical
simulation, starting from a disk with initial surface density $\Sigma(r)=1\,000$~g/cm$^2$ at 1~au, and exhibits a double gap around the orbits of 
Jupiter and Saturn. We assume that this surface density decays uniformly over time as $\exp(-t/\tau_{gas})$ with $\tau_{gas}$ = 5 My.
The aspect ratio of the disk is assumed to be:
 \begin{equation}
 h=\frac{H(r)}{r} =0.033r^{0.25}\ ,
 \end{equation}
 where $H(r)$ is the pressure scale height at radius $r$.

In this work, our preliminary numerical test showed that the GDF force is not more effective than aerodynamic gas drag. We decided nevertheless to keep both the gas drag and the GDF effect for completeness.

We used the same version of the N-body REBOUND code \citep{Rein2012,ReinSpiegel2015,ReinTamayo2015} modified by \citet{RIBEIRO2020113605} to perform the N-body simulations including the effects cited above as additional forces. 
In our simulations, the planetesimals were assumed to be non-interacting with each other.

\subsection{Evolution of planets and planetary embryos}
\label{inv}

The actual migration history of Jupiter and Saturn is not well known. Some works claimed that a Jupiter ending at 5 au could only have formed from a seed initially located beyond 20 au 
\citep{Bitshetal2015}. \citet{RIBEIRO2020113605} investigated the role of Jupiter migration in the planetesimal disk evolution and concluded that a long-range inward migration of Jupiter from 15 au or beyond 
would over-excite the inclinations of Kuiper belt objects with respect to what is presently observed. 
Thus, we adopt Jupiter at 5 au in a fixed orbit. A short range migration of Jupiter should not affect significantly the results.

In this work, we follow the \citet{Izidoroetal2015} model for the origin of Uranus and Neptune, where in this model 
Jupiter and Saturn are in a 3:2 resonance between Jupiter and Saturn \citep{NesvornyMorby2012}. However, there are other planetary configurations that have been considered in the recent literature. 
For example, \citet{2021Icarclem} (a) and (b) showed that a primordial 2:1 resonance between Jupiter and Saturn is advantageous in terms of its ability to consistently 
replicate the current Jupiter and Saturn eccentricities and the possibility that the giant planets’ primordial orbits were eccentric.

Because the simulation of \citet{Izidoroetal2015} has Jupiter at 3.5 au, we rescaled the semimajor axes of all bodies by a factor of $5/3.5$. 
The radial surface density profile of the gas is also rescaled to keep the gaps centered on the giant planets' orbits. More specifically, given 
the rescale factor $R(t)$ on semimajor axes, the surface density at the heliocentric distance r is computed as: 
 \begin{equation}
\frac{\Sigma\left(\frac{r}{R(t)}\right)}{(R(t))^{2}}, 
 \end{equation}
\noindent where $\Sigma$ is the surface gas density.

\begin{figure*}
\gridline{\fig{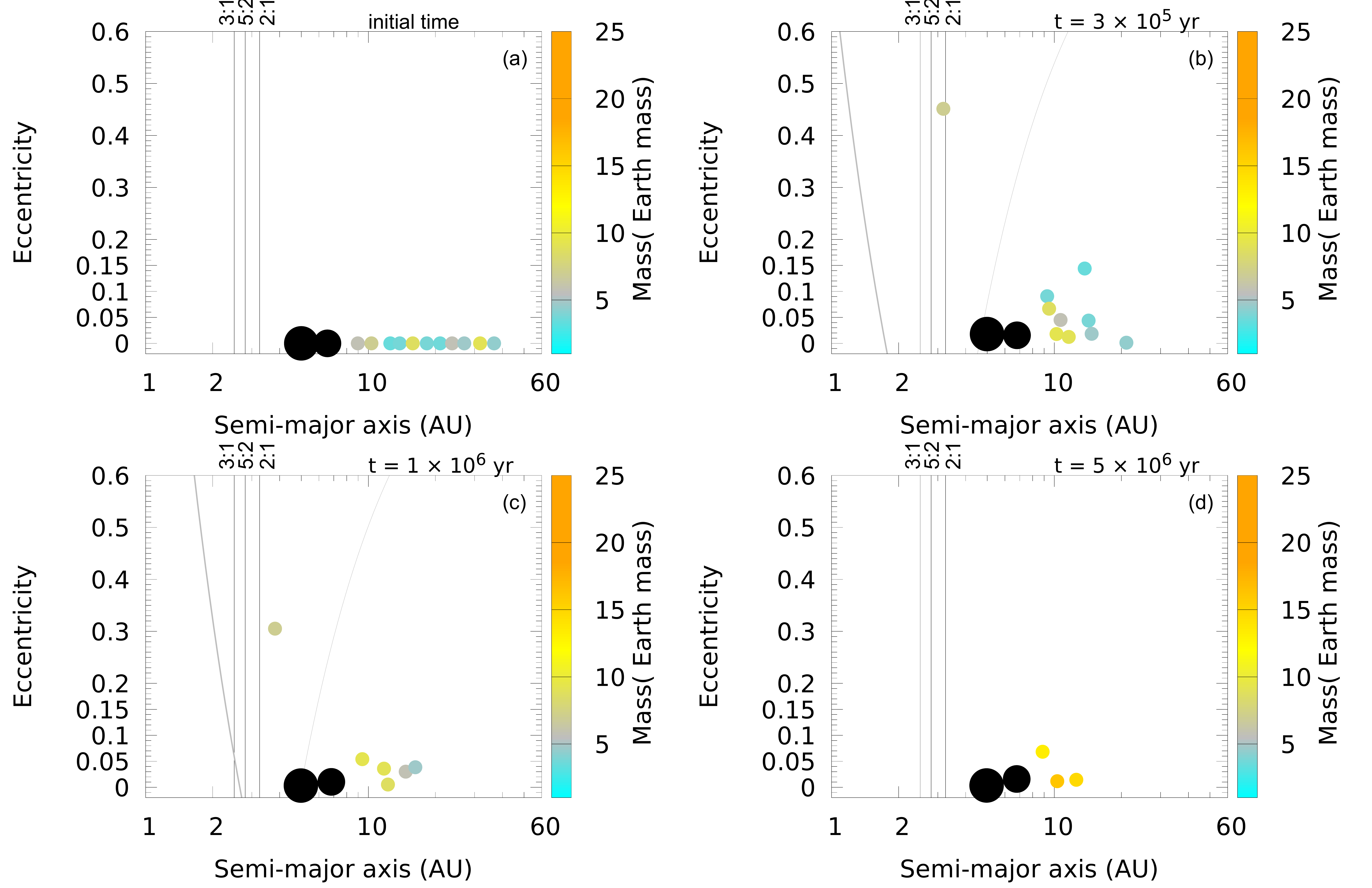} {1\textwidth}{}
          }
 \gridline{\fig{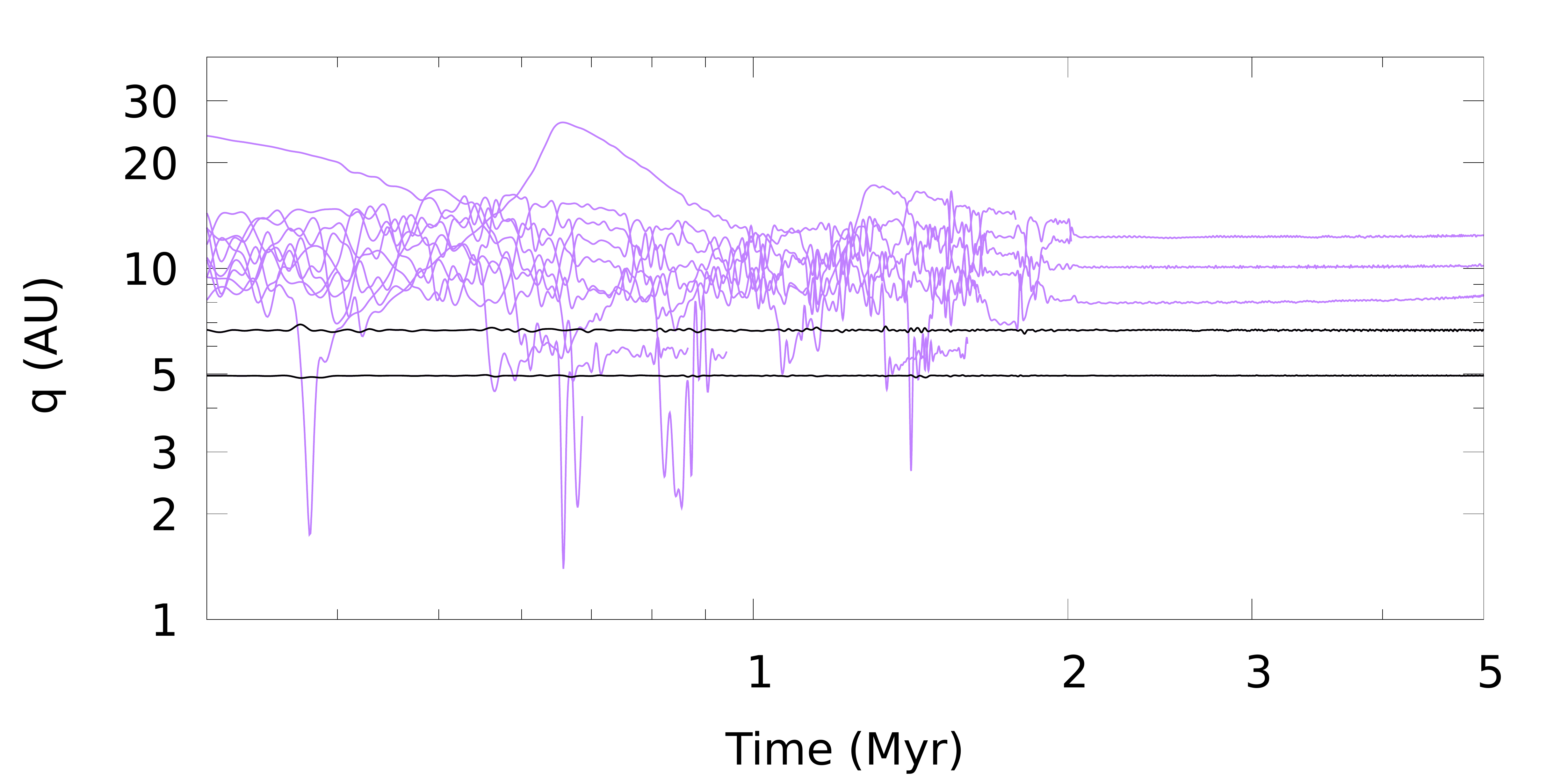} {0.6\textwidth}{(e)}}

\caption{Panels (a) - (d) are semimajor axis vs. eccentricity snapshots of the system with Jupiter, Saturn and a collection of protoplanets of the considered \citet{Izidoroetal2015} simulation. 
The 3:1, 5:2 and 2:1 mean motion resonance with Jupiter are presented by vertical lines. The gray lines represent the boundaries in (a,e) 
space within which planetesimals are intersected by an invader. Panel (e) represents a complete evolution of planets (black lines) and protoplanets (magenta lines) pericenters in the same simulation. \label{fig:interpolation1}}
\end{figure*}

Figure \ref{fig:interpolation1} illustrates our reproduction of the \citet{Izidoroetal2015} simulation using the interpolation scheme and scaling the 
initial semimajor axes. We show in panel (a) the initial semimajor axis and eccentricities of the planetary system (all initially zero) which contains 
Jupiter (at $5.0$~au), Saturn (at $6.55$~au), and a collection of $11$ protoplanets distributed with semimajor axis in a range from $8$ to $40$~au with masses between $3$ to $9$ $M_\oplus$.

In Panels (b) and (c), we show snapshots of the semimajor axis and eccentricity of Jupiter, Saturn, and the protoplanets at two different times: $1 \times 10^{5}$~yr and $1 \times 10^{6}$~yr 
respectively. At these times we can see that most protoplanets migrated inward, although they were blocked by Jupiter and Saturn remaining beyond the giant planets' orbits. Most of these protoplanets will 
grow more massive through mergers. However some protoplanets can enter the inner region of the Solar System ($a < 5$~au) for a short period of time due to scattering by Jupiter. 
We define these protoplanets as \textit{invaders}.

The \textit{invaders} are particularly interesting objects because they may affect the planetesimals capture process in the inner Solar System. 
The gray lines in the Panels (b) and (c) of Figure  \ref{fig:interpolation1} show the boundary in (a,e) space beyond which the planetesimals are intersected by the \textit{invader}'s trajectory. 
These lines cross the mean motion resonances (MMRs) with Jupiter (the 3:1, 5:2, and 2:1 MMRs) are presented in panels (b) and (c) of Figure  \ref{fig:interpolation1} by vertical black lines.
In Panel (d) of Figure  \ref{fig:interpolation1}, we show the final semimajor axis and eccentricity of Jupiter, Saturn, Uranus, Neptune and of a fifth comparable-mass planet at $5.1$~Myr, the end of the protoplanetary gas disk lifetime. 

Observing the pericenter evolution of the protoplanets, showed in Panel (e) of Figure  \ref{fig:interpolation1}, we note that the \textit{invaders} enter deeply into the inner Solar System, 
down to a distance between 1.3~au to 3~au during the simulation time. These \textit{invaders} can potentially have close encounters with the planetesimals in the inner Solar System.

\subsection{Distribution of planetesimals}\label{sec:dist}

The large size of Ceres suggests that it has not been altered significantly 
during its collisional evolution \citep{MccordSotin2005}. Thus, we assume that Ceres formed in the trans-Saturnian region and
dynamically evolved keeping its current size and density. The size and density of
Ceres are very important due to the effects of the gas disk.

We used $1,000$~km sized planetesimals to build a disk distributed beyond
the orbits of the giant planets, throughout the region occupied by the
protoplanets (8~au to 40~au). We divided this region into three different
reservoirs constrained by chemical models that explain the surface colors
of observable Kuiper Belt objects \citep{Brownetal2011}.

\citet{Brownetal2011} suggested that planetesimals are characterized by strong heliocentric-distance gradients in surface composition due to ultraviolet and particle irradiation. This process
is called volatile evaporation and could explain the surface colors that are seen today in the objects of the Kuiper Belt. Based on this model, objects as large as Ceres can always retain on its surface H$_{2}$O, CO$_{2}$ and H$_{2}$S.
Between 20-30 au the planetesimals can retain also: HCN, C$_{2}$H$_{6}$, CH$_{3}$OH, while more distant planetesimals (from 30-50 au) can also have NH$_{3}$.

\begin{figure}[h]
    \begin{center}
    \includegraphics[scale=0.35]{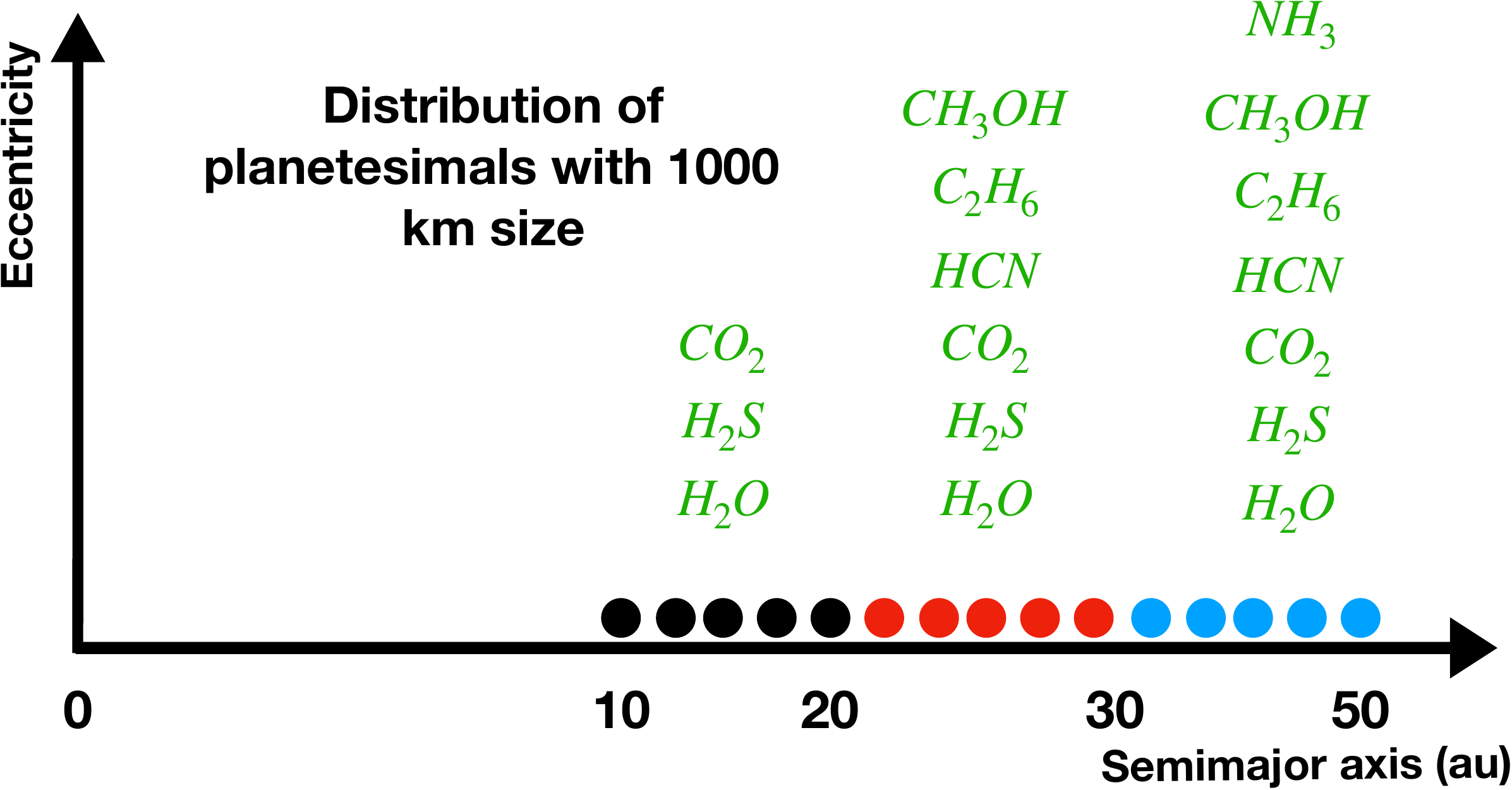}
  \caption{Initial distribution for planetesimals of 1\,000~km size in our simulations. The planetesimal are represented by black, red and blue circles. The text in green colors report the molecular species that are expected to survive on 
  the surface of the planetesimals in these regions. Adapted by \citet{Brownetal2011}.}
   \label{fig:traject}
    \end{center}
    \end{figure}

Adapted from Figure 1 of \citet{Brownetal2011}, Figure \ref{fig:traject} shows our initial distribution of planetesimals. The planetesimals are differentiated by color (black, red, and blue) based on their initial heliocentric
distances and the reservoir of molecular species that are expected to be present at their surface (green colored text). 

Ceres has ammonia on its surface but this is not an indication that Ceres could have formed only in the blue region. Moreover, Ceres spent at least almost 4.5~Gy in the
asteroid belt, so it \emph{should} have lost by irraditation its surface ammonia, regardless of where it formed. It may be possible that ammonia is rejuvenated by outgassing from the interior or it 
reacted chemically to form more refractory
ammoniated salts. In any of these cases, if this happened in the asteroid belt
it could have happened anywhere in the disk. So, Ceres does not need to have been
formed beyond 30 au. Instead, to accrete ammonia in the first place, Ceres should have formed
beyond the ammonia snowline ($\sim90$K). In an early disk \citep{Bitshetal2015} this
could be 7-10 au. In a late disk the ammonia snowline can go into the asteroid
belt, but at that time Jupiter was already formed and blocking the drift of
ammonia-rich dust, so  it is safe to assume that Ceres formed beyond Jupiter and
maybe even beyond Saturn. In summary, Ceres can be any object from the black, red or even blue region in Figure \ref{fig:traject}.

As initial conditions, we assume that in the black and red regions there are 5\,000 planetesimals each, in the blue region there are 10\,000 planetesimals.
Notice that these numbers are not related to the actual number of Ceres-size objects expected in the disk. A large number of objects is considered to 
have reliable statistical results on the probability of implantation in the asteroid belt. The resulting probabilities are then mutiplied by 
the number of Ceres-size objects provided by the size distribution of the original planetesimal disk of \citet{Nesvorny2016a}. The latter was determined from constraints
on the Kuiper belt structure and apply to the planetesimal disk in the 20-30 au region after the formation of Uranus and Neptune. We assume here for simplicity
that the same size distribution applies in the 10-20 au region and in the 30-40 au region namely in each region of Fig. \ref{fig:traject}.
In particular in this work we do not just check that the implantation of a Ceres-size object from the trans-Saturnian disk into the asteroid belt is possible, but we also 
wish to verify that the implantation of 1 object is consistent with the number of Ceres-size objects expected in the trans-Saturnian disks. We acknowledge that 
the latter is quite uncertain, given that we extrapolate the \citet{Nesvorny2016a} results to the whole disk. 
We reiterate that the planetesimals are non-interacting particles. The eccentricities and inclinations of the planetesimals
are initially chosen to be $10^{-3}$. Their argument of pericenter and longitude of ascending node are randomly initialized between 0$^{\circ}$ and 360$^{\circ}$.

 \subsection{Ceres Candidates Runs}
 \label{ceresrun}

We monitor the orbital evolution of each planetesimal and record the orbital state of planetesimals that are scattered inward (reaching temporarily $ 2 < a < 3.5$~au).  
We call these planetesimals \textit{Ceres candidates} and determine the fraction of planetesimals they represent from each source reservoir (see Section \ref{blacksc}).
Then, from the analysis of the dynamical evolution of Ceres candidates, we study the mechanisms (potential scattering off one of the invaders, resonances with the giant planets and gas drag) 
that may insert these planetesimals in stable orbits close to the current Ceres' orbit.
Many works have shown that most of the Solar System is chaotic \citep{LASKAR1990266,Lecar2001ChaosIT,K2015}. The influence of the \textit{invaders} make the inner Solar System even more chaotic.
Therefore, we make clones of the \textit{Ceres candidates} to measure the probability that the temporary capture is turned into a permanent capture by some irreversible process (e.g. gas drag, GDF or the encounter with an \textit{invader}).
To obtain the implantation statistics, we multiply the fraction of planetesimals that become Ceres candidates 
by the fraction of clones that are permanently trapped in the inner Solar System. The cloned bodies were given a small change in the original velocity of the parent body (fractionally changed by $10^{-6}$), where a similar procedure was used 
in \citet{BrasserandMorbidelli2013}.

\section{Results}
\label{results}

In this section we present results of a simulation from which we assess the number and the dynamical state of planetesimals that become \textit{Ceres candidates}. 
The initial setup of our numerical simulations and the prescribed interactions are described in Section \ref{Methods}.

\subsection{Fate of Planetesimals}
\label{blacksc}

\begin{figure*}
\gridline{\fig{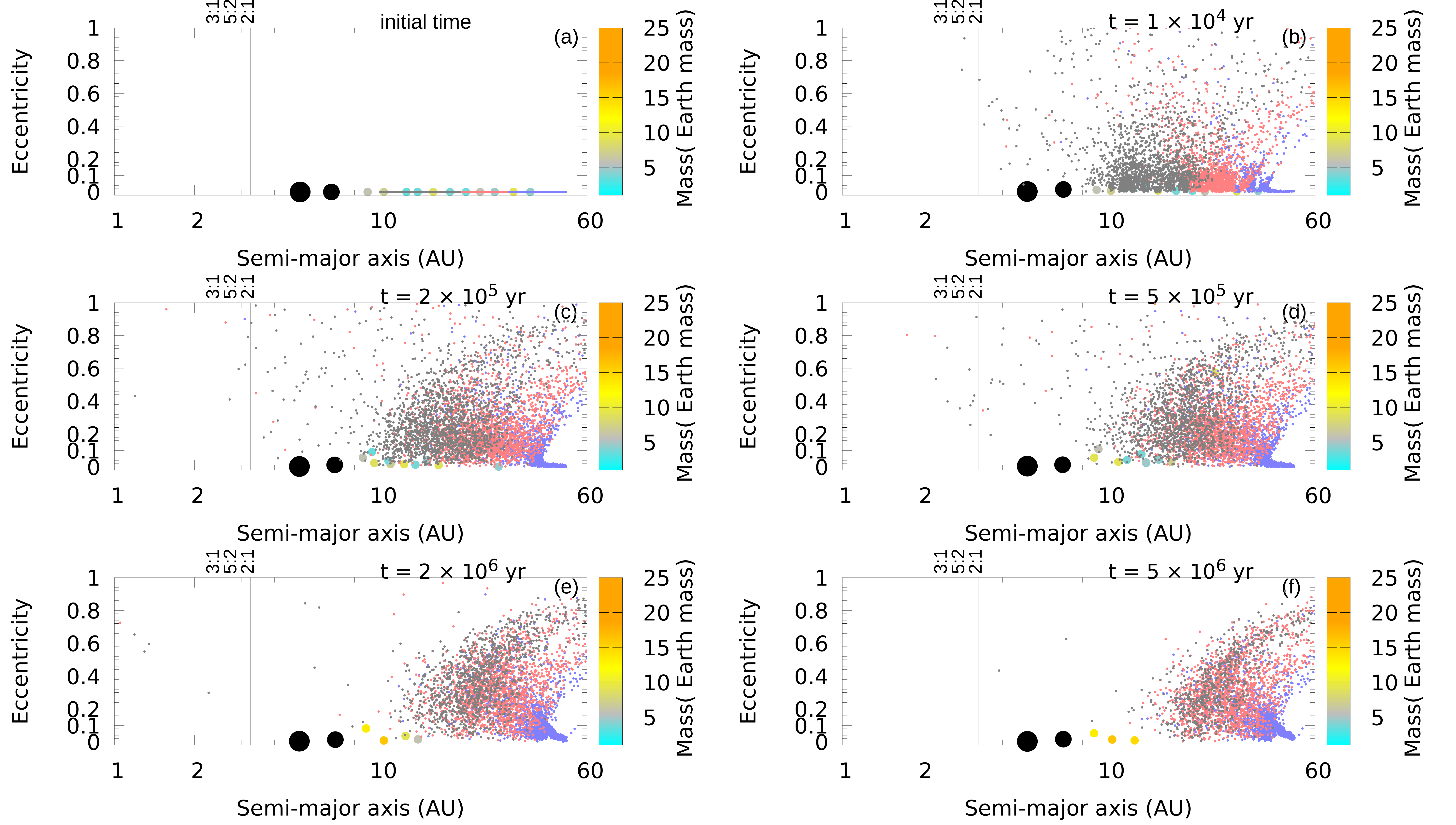} {1.11\textwidth}{}}

\caption{Panels (a) - (f) show eccentricity/semimajor  axis snapshots of the dynamical evolution of the planetesimals. The color of a planetesimal is associated to its initial location in the black, red, or 
blue disk regions. The color of the protoplanets is associated to the color scale representing the mass. Jupiter and Saturn are in black. The vertical lines are 3:1, 5:2 and 2:1 MMR with Jupiter.
\label{fig:black1}}
\end{figure*}

\begin{figure*}
\gridline{\fig{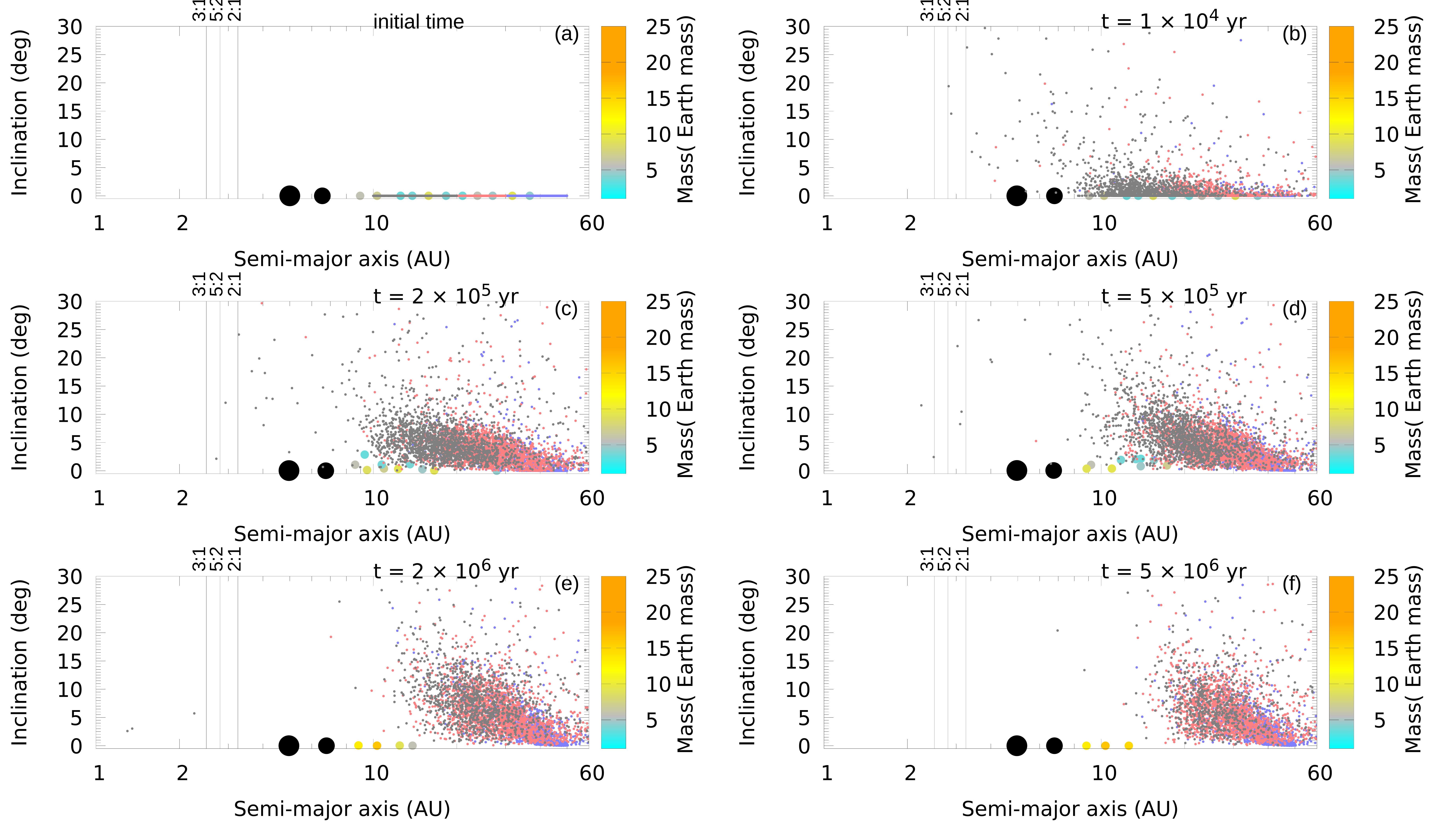} {1.11\textwidth}{}}

\caption{ Same as Fig. 3, but illustrating the evolution in semi major axis vs. inclination.
\label{fig:black2}}
\end{figure*}

Figures \ref{fig:black1} and \ref{fig:black2} show snapshots of the simulation where planetesimals were initialized in the black, red, and blue
regions. The sculpting process of the planetesimal disk is detailed in
\citet{RIBEIRO2020113605}. During the dynamical evolution of the planetary
embryos, the planetesimal population is depleted by close encounters and collisions with
protoplanets or either Jupiter or Saturn.

A large fraction of the planetesimal population persists in the outer region of the Solar
System (a $>$ 20~au). Only a small fraction of the planetesimal population is scattered inward by close encounters with the planets and reaches the inner Solar System. The orbits
of some of these planetesimals have a semimajor axis within the asteroid-belt region but eccentricities that are large enough to bring them close to Jupiter's trajectory. 
Therefore, they are not on permanent stable
orbits (see Section \ref{ceresmecha} to see the mechanisms to get permanent stable
orbits). 

A higher fraction of the planetesimal population starting in the black region becomes \textit{Ceres candidates}, compared to the planetesimal populations starting in either of the two other reservoirs (See Table \ref{tab:mathmode0}). 
These fractions are 11.11\%, 4.26\% and 0.28\% for the black, red, and blue populations respectively. The planetesimals in the red and blue regions need to reach the black region via close encounters with the embryos, 
before having a chance to be sent into the inner Solar System by encounters with Saturn and Jupiter. This explains their smaller implantation probability compared to the objects native from the black portion of the disk.

\begin{deluxetable*}{ccCrlc}[h!]
\tablecaption{\label{tab:mathmode0}}
\tablecolumns{3}
\tablenum{1}
\tablewidth{0pt}
\tablehead{
\colhead{} &
\colhead{Black region} &
\colhead{Red region} &
\colhead{Blue region} 
}
\startdata
\textbf{$N_{total}$} & $5,000$ & $5,000$ & $10,000$  \\
\textbf{$N_{temp}$} & 587 &  213 & 28 \\
\textbf{$C_{EM}$} & $0.111$ & $0.0426$ & $0.0028$ \\
\enddata
\caption{Statistics of planetesimals that temporarily evolved into the inner Solar System ($a< 3.5$ au). $N_{total}$ is the number of initial of planetesimals for each region. $N_{temp}$ and $C_{EM}$ 
are the number and fraction ($\frac{N_{temp}}{N_{total}}$) of planetesimals that became \textit{Ceres candidates} respectively.}
\end{deluxetable*}

\begin{deluxetable*}{ccCrlc}[h!]
\tablecaption{\label{tab:mathmode}}
\tablecolumns{2}
\tablenum{2}
\tablewidth{0pt}
\tablehead{
\colhead{} &
\colhead{Stable orbits} 
}
\startdata
\textbf{$N_{eff}$} & 294,087   \\
\textbf{$N_{ab}$} & 3012  \\
\textbf{$N_{inner}$} & 800  \\
\textbf{$N_{central}$} & 897 \\
\textbf{$N_{outer}$} & 1315 \\
\enddata
\caption{Statistics of planetesimals implanted in orbits in the asteroid belt. $N_{eff}$ is the total effective number of planetesimals (number of Ceres candidates times the number of clones for each candidate)
represented in our simulations. $N_{ab}$ is the number of planetesimals implanted in the asteroid belt region (2.1 $<$ a $<$ 3.25 au, e $<$ 0.4 and i $<$ 40 degrees).
We used a broad asteroid belt region because some high-e or high-i objects can then be pushed within the current asteroid belt boundary during the subsequent dynamical instability 
of the giant planets \citep{deiennoet2016}. 
$N_{inner}$, $N_{central}$ and $N_{outer}$ are the number of planetesimals implanted in the inner asteroid belt region ($2.1 < a < 2.5$ au), central asteroid belt region
between the 3:1 MMR and 5:2 MMR ($2.5 < a < 2.82$ au) and outer main belt region ($2.82 < a < 3.25$ au) between the 5:2 MMR and 2:1 MMR, respectively.}
\end{deluxetable*}

    \begin{deluxetable*}{ccCrlc}[h!]
\tablecaption{\label{tab:mathmode3}}
\tablecolumns{3}
\tablenum{3}
\tablewidth{0pt}
\tablehead{
\colhead{} &
\colhead{Black region} &
\colhead{Red region} &
\colhead{Blue region} 
}
\startdata
\textbf{Asteroid belt region} & $1.20 \times 10 ^{-3}$ & $4.36 \times 10 ^{-4}$ & $2.87 \times 10 ^{-5}$  \\
\textbf{Inner Main belt region} & $3.19 \times 10 ^{-4}$ &  $1.16 \times 10 ^{-4}$ & $7.62 \times 10 ^{-6}$ \\
\textbf{Central Asteroid belt region} & $3.58 \times 10 ^{-4}$ &  $1.30 \times 10 ^{-4}$ & $8.54 \times 10 ^{-6}$ \\
\textbf{Outer Asteroid belt region} & $5.25 \times 10 ^{-4}$ &  $1.90 \times 10 ^{-4}$ & $1.25 \times 10 ^{-5}$ \\
\enddata
\caption{The capture efficiency in the asteroid belt per \textit{Ceres candidate}. Here, the asteroid belt region is the region with 2.1 $<$ a $<$ 3.25 au, e $<$ 0.4 and i $<$ 40 degrees. The inner, central and outer asteroid 
belt regions are respectively the regions with $2.1 < a < 2.5$ au, $2.5 < a < 2.82$ au and $2.82 < a < 3.25$ au, e $<$ 0.4 and i $<$ 40 degrees.}
\end{deluxetable*}

    \begin{deluxetable*}{ccCrlc}[h!]
\tablecaption{\label{tab:mathmode4}}
\tablecolumns{3}
\tablenum{4}
\tablewidth{0pt}
\tablehead{
\colhead{} &
\colhead{Black region} &
\colhead{Red region} &
\colhead{Blue region} 
}
\startdata
\textbf{Asteroid belt region} &4-17 & 1-6 & 0.10-0.40 \\
\textbf{Inner Main belt region} &1-4 & 0.41-1 & 0.02-0.10\\
\textbf{Central Asteroid belt region} &1-5 & 0.46-2 & 0.03-0.11\\
\textbf{Outer Asteroid belt region} &2-7 & 0.68-2 & 0.04-0.17\\
\enddata
\caption{Probability that a Ceres-size planetesimal is implanted in the asteroid belt in our scenario. When the probability is larger than 1, more
than one implanted object is expected.}
\end{deluxetable*}

\begin{figure*}
\gridline{\fig{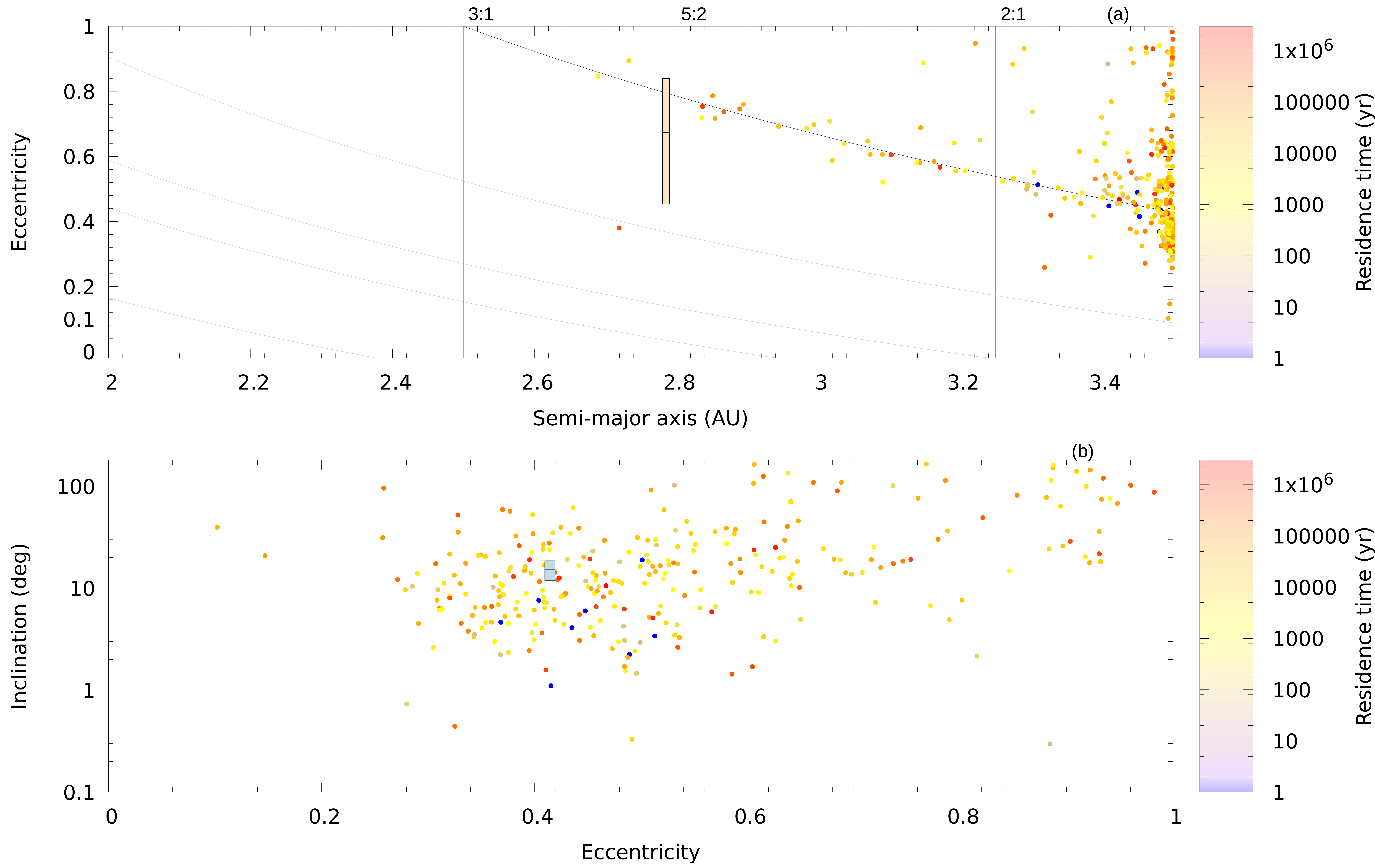} {1\textwidth}{}}
     
\caption{Panels (a) and (b) show, for the planetesimals started in the black region, the orbital elements achieved when they are first recorded as \textit{Ceres candidates}. The
  color represents the time that each planetesimal subsequently spent as a \textit{Ceres candidate}. The yellow and blue boxes show respectively the eccentricity and inclination evolution 
  of a single \textit{Ceres candidate}. The black and gray curve represent the boundary in (a,e) space above which a planetesimal crosses Jupiter's orbit and crosses four different \textit{invaders}' orbits, respectively.  
\label{fig:black4}}
\end{figure*}
        
Figure \ref{fig:black4} shows the osculating elements (semimajor  axis, eccentricity, and inclination) of the planetesimals when they first become \textit{Ceres candidates}. 
Note that there is a pile up of objects near 3.5 au. This is mostly due to a sampling bias, given that the bodies are plotted when they first enter in the \textit{Ceres candidate} region, whose outer 
boundary is 3.5 au.
Figure \ref{fig:black4} (a) shows that most planetesimals are in/close to the scattering region of Jupiter, where the black curve represents the boundary in (a,e) space above
which a planetesimal crosses Jupiter's orbit. Gray curves represent the boundary in (a,e) space above which a planetesimal crosses four different \textit{invaders}' orbits. Many of 
the planetesimals spend less than 500 thousand years as a \textit{Ceres candidate}. However, there are a few exceptional planetesimals that spend 1 Myr 
or more with a semimajor axis in the asteroid belt region. Therefore, there are two populations of \textit{Ceres candidates}:
those which remain in the scattering regime (and are therefore short-lived) and those who manage to decouple from Jupiter due to resonance capture, encounters with invaders, or due to gas drag. 

Figure \ref{fig:black4} (b) shows that the inclinations of \textit{Ceres candidates} have a very excited distribution, 
with some planetesimals even presenting a retrograde orbit (inclinations larger than 90$^{\circ}$).

We did not find any correlation between the orbits trapped in mean motion resonances (MMR) with Jupiter and their initial location in the trans-Saturnian disk. This is a consequence of the chaotic evolution of the system.
For the same reason, the distribution of the osculating elements of the planetesimals from the red and blue regions of the disk 
are very similar to the those of the planetesimals originating from the black region, although their number is much smaller.

\subsection{Dynamical mechanisms for the implantation of Ceres-sized planetesimal}
\label{ceresmecha}

\begin{figure*}
\gridline{\fig{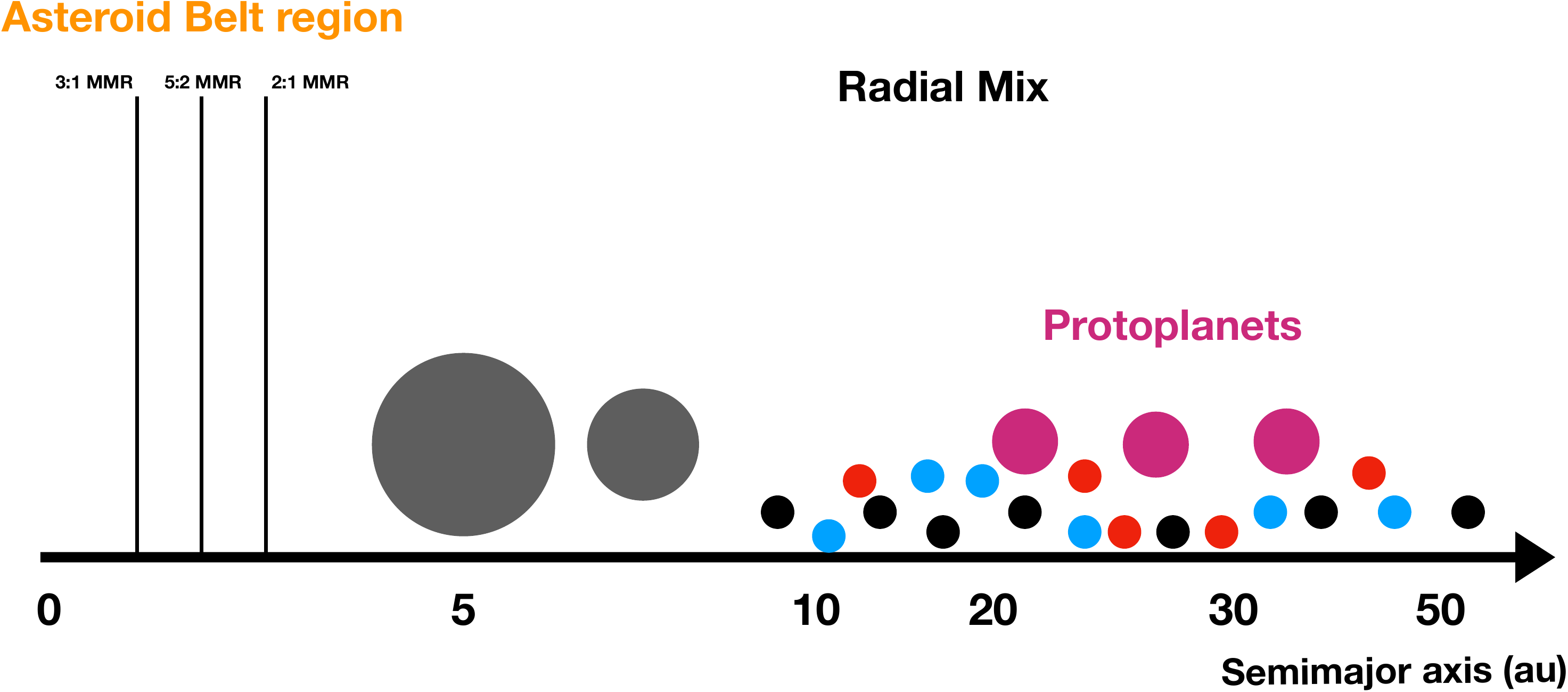} {0.5\textwidth}{(a) Phase (i)}
\fig{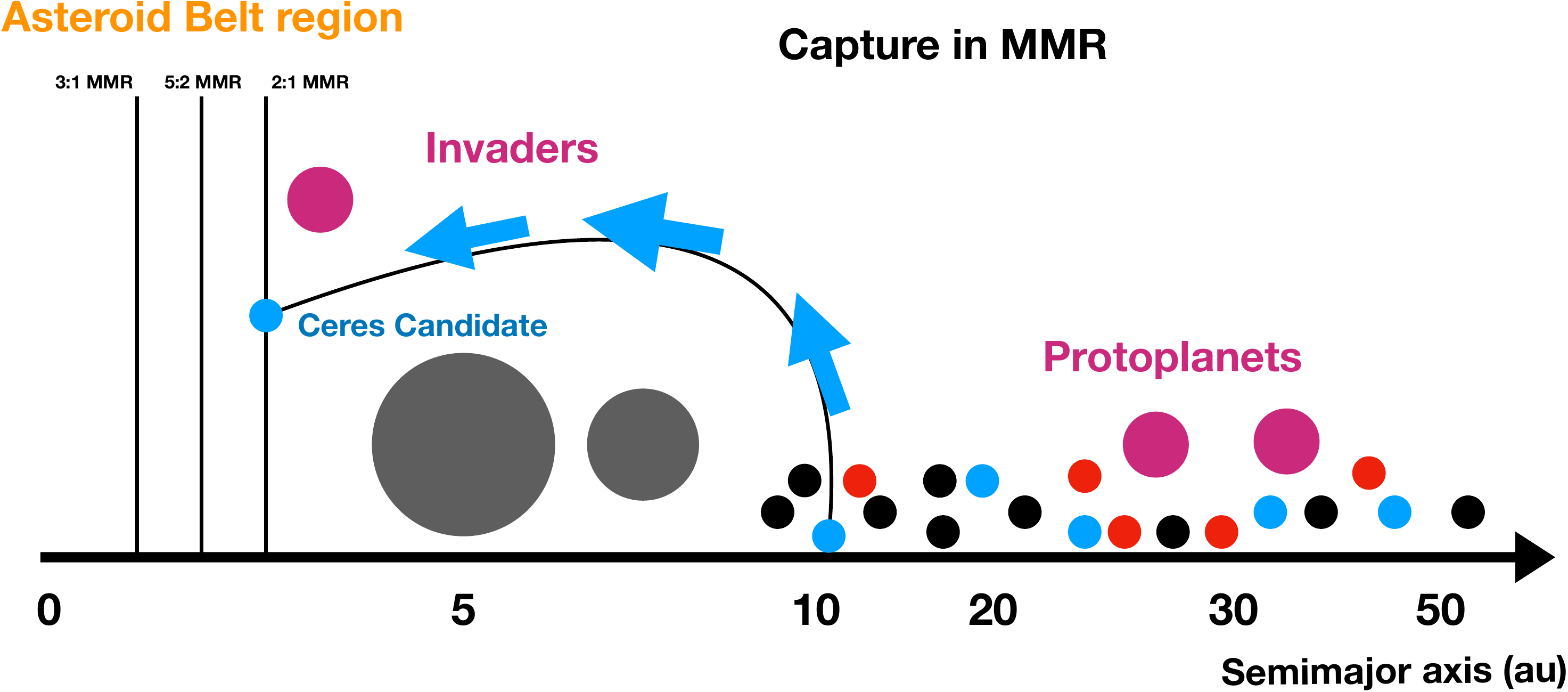} {0.5\textwidth}{(b) Phase (ii)}
          }
          
          \gridline{\fig{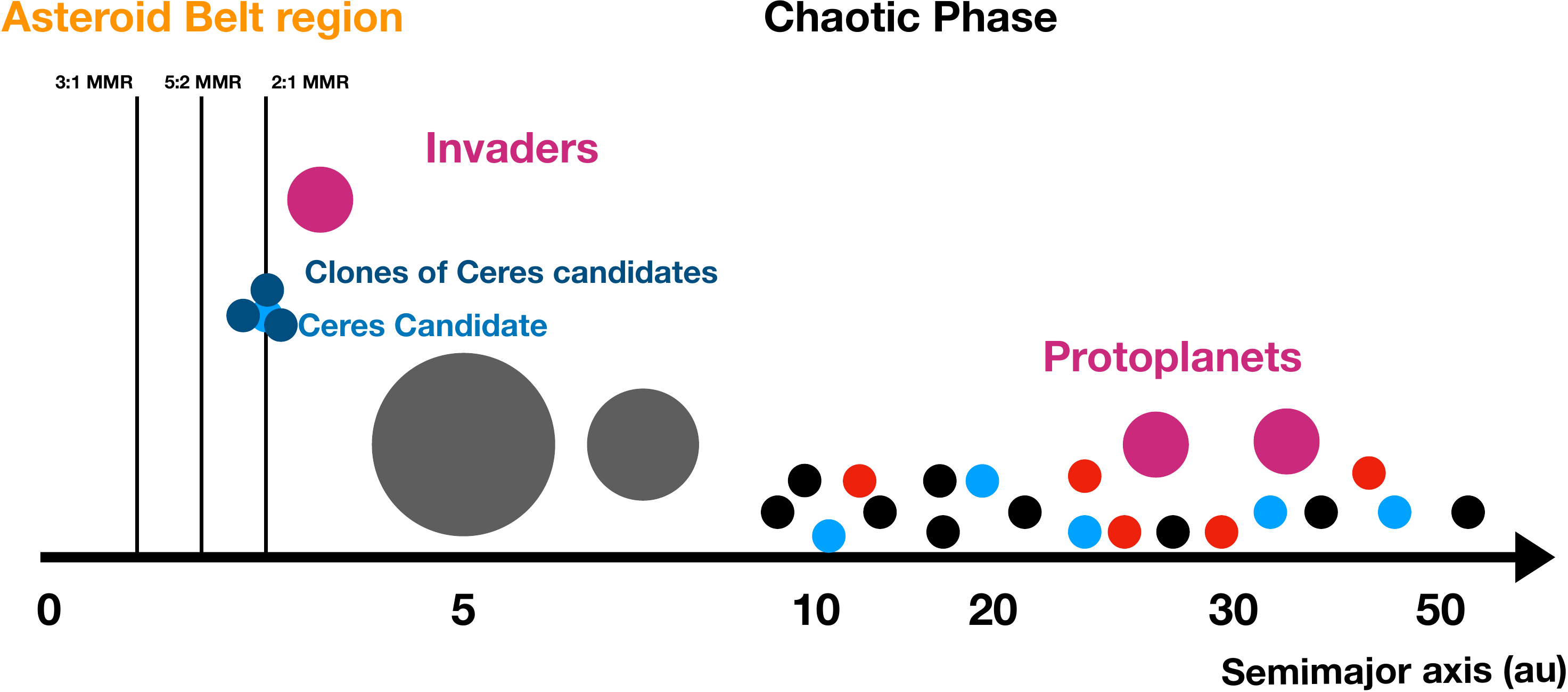} {0.5\textwidth}{(c) Phase (iii)}
\fig{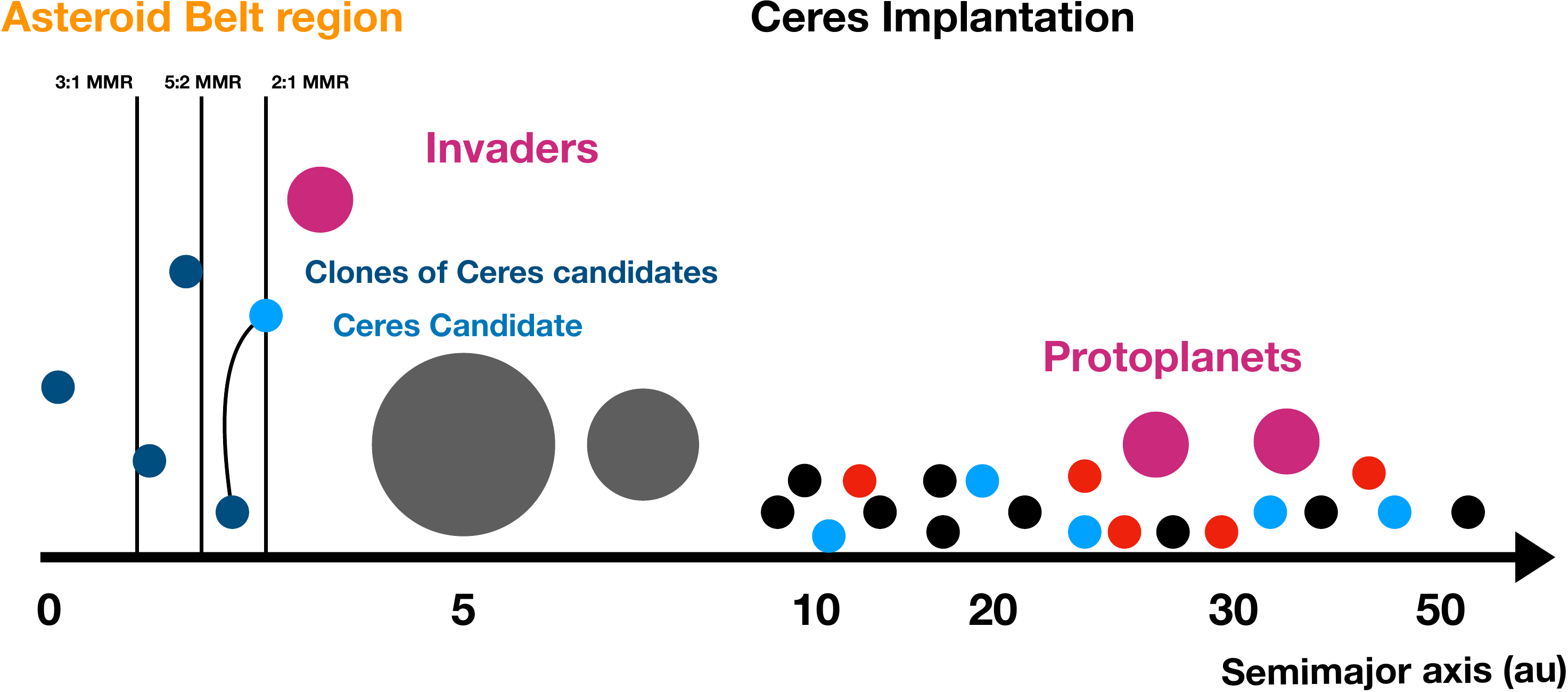} {0.5\textwidth}{(d) Phase (vi)}
          }
          
\caption{Sketch of the different steps of the scenario envisioned to implant a Ceres-sized planetesimals in the asteroid belt (see text to more details).}

\label{fig:black45}
\end{figure*}

We explore a scenario to implant a Ceres-sized planetesimals stably near the real Ceres' orbit from the outer region of the Solar System. A scheme of the scenario is presented in 
Figure \ref{fig:black45}.
The scenario starts with a fast radial mixing in the position of the planetesimals in the outer planetesimal disk (Fig. \ref{fig:black45} (a)). The radial mixing is caused by
a fast scattering process of the protoplanets in the first thousand years (Fig. \ref{fig:black1} (b)).
Once close to Saturn (even if originally coming from the outer disk as the blue planetesimal on which Fig. \ref{fig:black45} (b) focuses on) a planetesimal
can be injected into the inner Solar System by encounters with the giant planets. 

When the scattered planetesimals become \textit{Ceres candidates} their orbital eccentricities are typically too large and their orbits
are not dynamically stable because they cross the orbits of the giant planets. However, three mechanisms can drop the eccentricities of the planetesimals, allowing 
them to avoid further close encounters with Jupiter: (i) mean motion resonances with Jupiter coupled with the Kozai resonance,
(ii) close encounters with some \textit{invaders}, and (iii) the effect of the gas drag.
We discuss each of these processes below. 

(i) Once a planetesimal is scattered into a mean motion resonance, it is phase-protected from 
further close-encounters with Jupiter for some time, where secular effects can drop the planetesimal's eccentricity. 
The Kozai resonance within a mean motion resonances is a particularly strong secular effect that can drop the eccentricities of planetesimals
with initially moderate to large inclinations \citep{2020RaymondNesvorny}. 

(ii) When a planetesimal encounters an \textit{invader} with a random phase, the planetesimal's eccentricity can either increase or decrease. If the eccentricity decreases the planetesimal can reach an orbit that does not come
close to Jupiter at aphelion, and which can become stable when the \textit{invader} is removed. 

(iii) Gas drag and gaseous dynamical friction are a functions of the density of the gaseous disk, the orbit and the size of the planetesimal.
In our numerical simulations, the GDF is too weak to drop the eccentricities of these objects. However, the aerodynamic gas drag can be sufficient to drop the eccentricities and the
inclinations of the planetesimals particularly once they become protected from planetary encounters (e.g. in a resonance with Jupiter or after the removal of the invaders).

\begin{figure}[h]
    \begin{center}
    \includegraphics[scale=0.3]{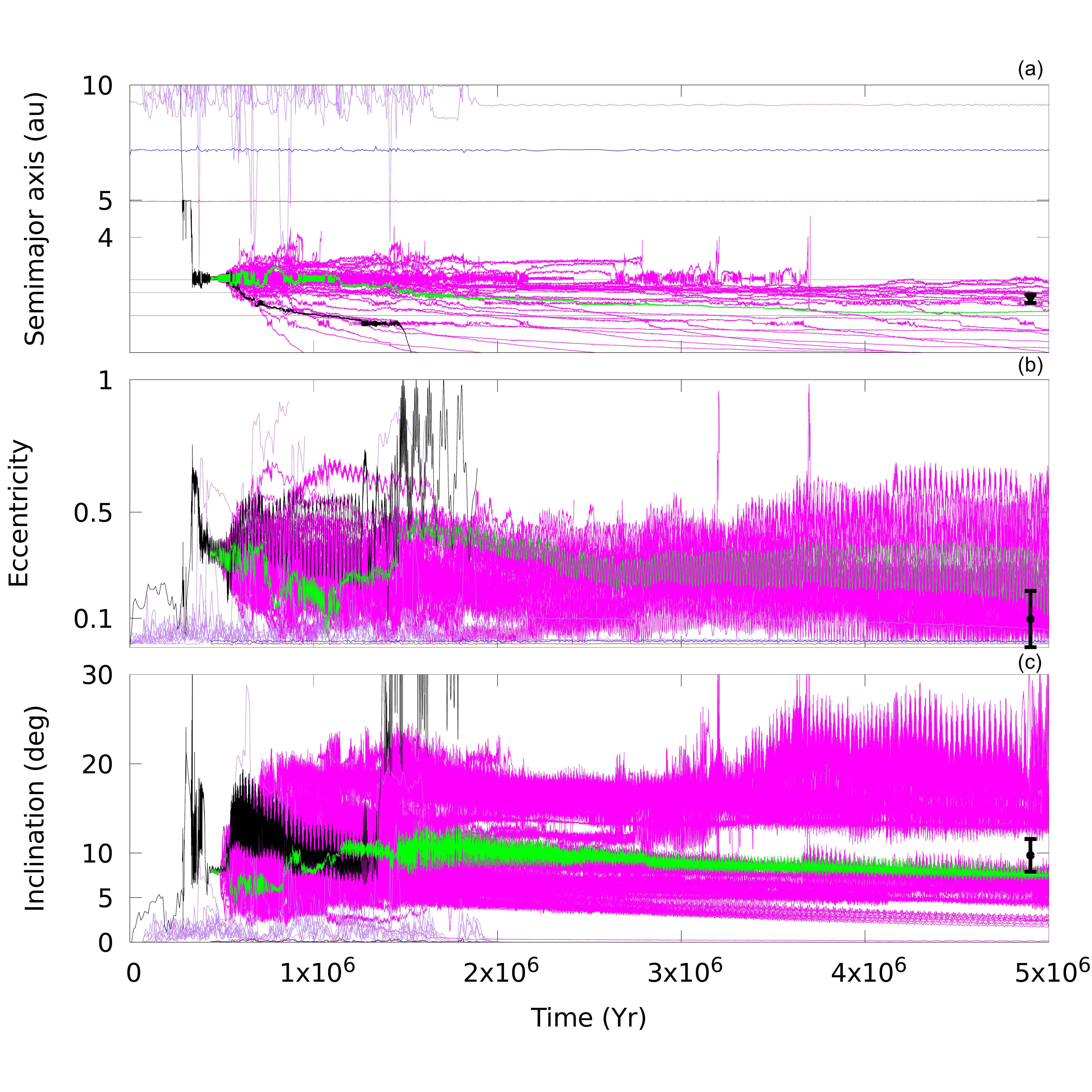}
  \caption{Panels (a), (b) and (c) show the evolution of the semimajor axis, eccentricity and inclination of Jupiter (red line), Saturn (blue line), protoplanets (purple lines), 
the original implanted planetesimal (black line) and of its clones (pink and green lines). The green line represents a clone planetesimal that reaches an orbit very close to the current orbit of Ceres (black point).
The error bars denote the range of eccentricity and inclination that Ceres spans over long timescales \citep{2019A&A...622A..95V}.}
   \label{fig:blackplant}
    \end{center}
    \end{figure}

These three mechanisms do not necessarily act disjointly. We have seen in Figure \ref{fig:blackplant}
that some planetesimals are placed in mean motion resonances with Jupiter where their dynamical lifetime is 
enhanced, giving more time to mechanisms (ii) and (iii) to act, in addition to secular effects. Thus, the second phase of our scenario is the capturing of planetesimals 
in mean motion resonance with Jupiter. 

The mechanisms (i)-(iii) discussed above can chaotically diffuse the orbits of Ceres-like candidates. Thus, chaotic diffusion is the last part of the scenario (Fig. \ref{fig:black45} (d))
to place a \textit{Ceres candidate} onto an orbit close enough to the current orbit of Ceres. Obviously, once the gas is removed from the protoplanetary disk, all \textit{invaders} are removed and Ceres-candidates 
are extracted from the original MMR, mechanisms (i)-(iii) 
stop and the capture on a Ceres-like orbit becomes permanent. 

To demonstrate the effectiveness of chaotic diffusion of the orbits of the \textit{Ceres' candidates}, we performed simulations cloning each \textit{Ceres candidate} 500 times. Figure \ref{fig:blackplant} shows the evolution of the semimajor axis, eccentricity and inclination of Jupiter,
Saturn (red and blue lines respectively), of the protoplanets (purple lines), of a \textit{Ceres candidate} (black line), and of 50 of its clones (pink lines). As a result of the 
chaotic diffusion the orbital evolution of the clones is different from that of the original \textit{Ceres candidate} placing some of them in orbits that reach a stable state in the end. 
In green color, we represent a planetesimal that reaches an orbit very close to the current orbit of Ceres (black star point in Figure \ref{fig:blackplant}). 

\begin{figure}[h]
    \begin{center}
    \includegraphics[scale=0.35]{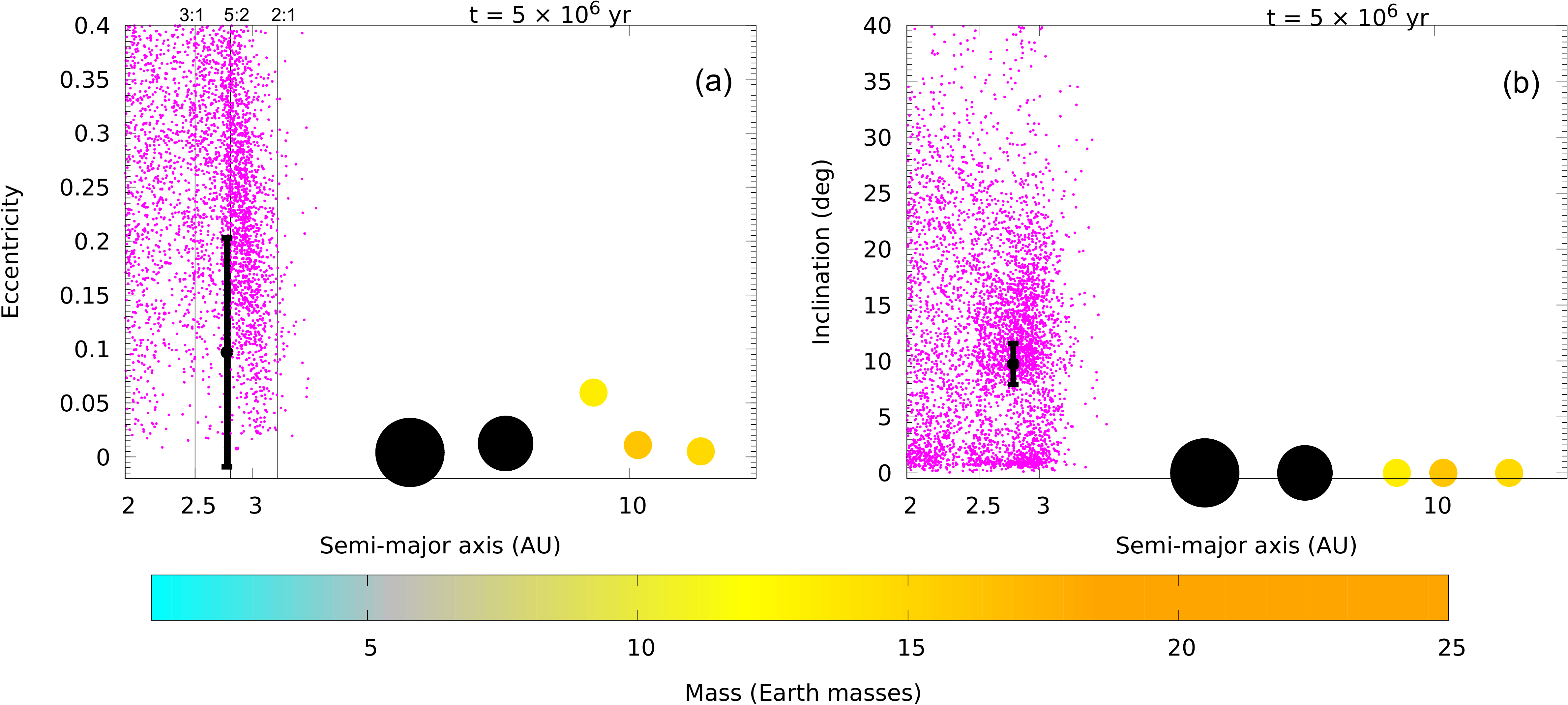}
  \caption{Panels (a) and (b) show the eccentricity, inclination and semimajor axis of the cloned planetesimals at the end of the gas phase. The error bars denote the range of eccentricity and inclination that Ceres spans over long timescales \citep{2019A&A...622A..95V}.}
   \label{fig:black120}
    \end{center}
    \end{figure}

In order to make a statistical analysis of the efficiency of capture of planetesimals in the asteroid belt region, we focus on the black planetesimals (planetesimals that started in the region between 10 to 20 au)
because they have more objects becoming Ceres candidates (about 587 planetesimals), but they are a representative also of the dynamical evolution for the red and blue planetesimals
once the latter start to be scattered by the giant planets. The cloning process gives us a total number of 294,087 planetesimals, which is equivalent to having initially a total of 2.5 million
planetesimals in the black portion of the disk for the purpose of statistics. Figures \ref{fig:black120} (a) and (b) show the eccentricity, inclination and semimajor axis of the giant planets and the 
cloned planetesimals (pink points), at the end of the gas disk phase. Note that many of them will not be stable on a long timescale because their eccentricities and inclinations exceed the boundaries of our extended asteroid
belt (e = 0.4, i = 40$^{\circ}$). Note that implanted planetesimals cover the whole asteroid belt.
Some planetesimals can reach orbits close to the current orbits of Ceres.  

\begin{figure*}
\gridline{\fig{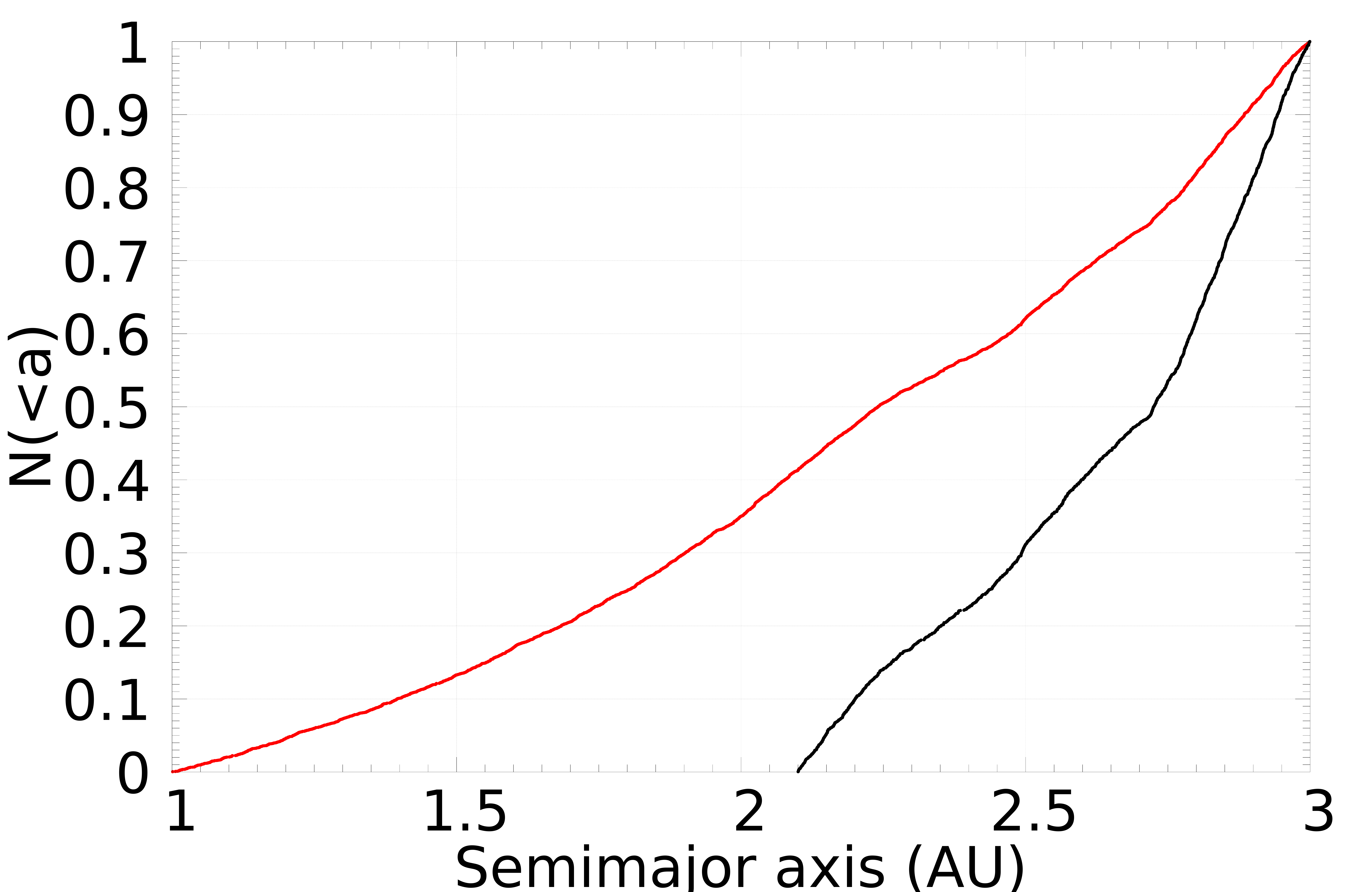} {0.3\textwidth}{(a)}
\fig{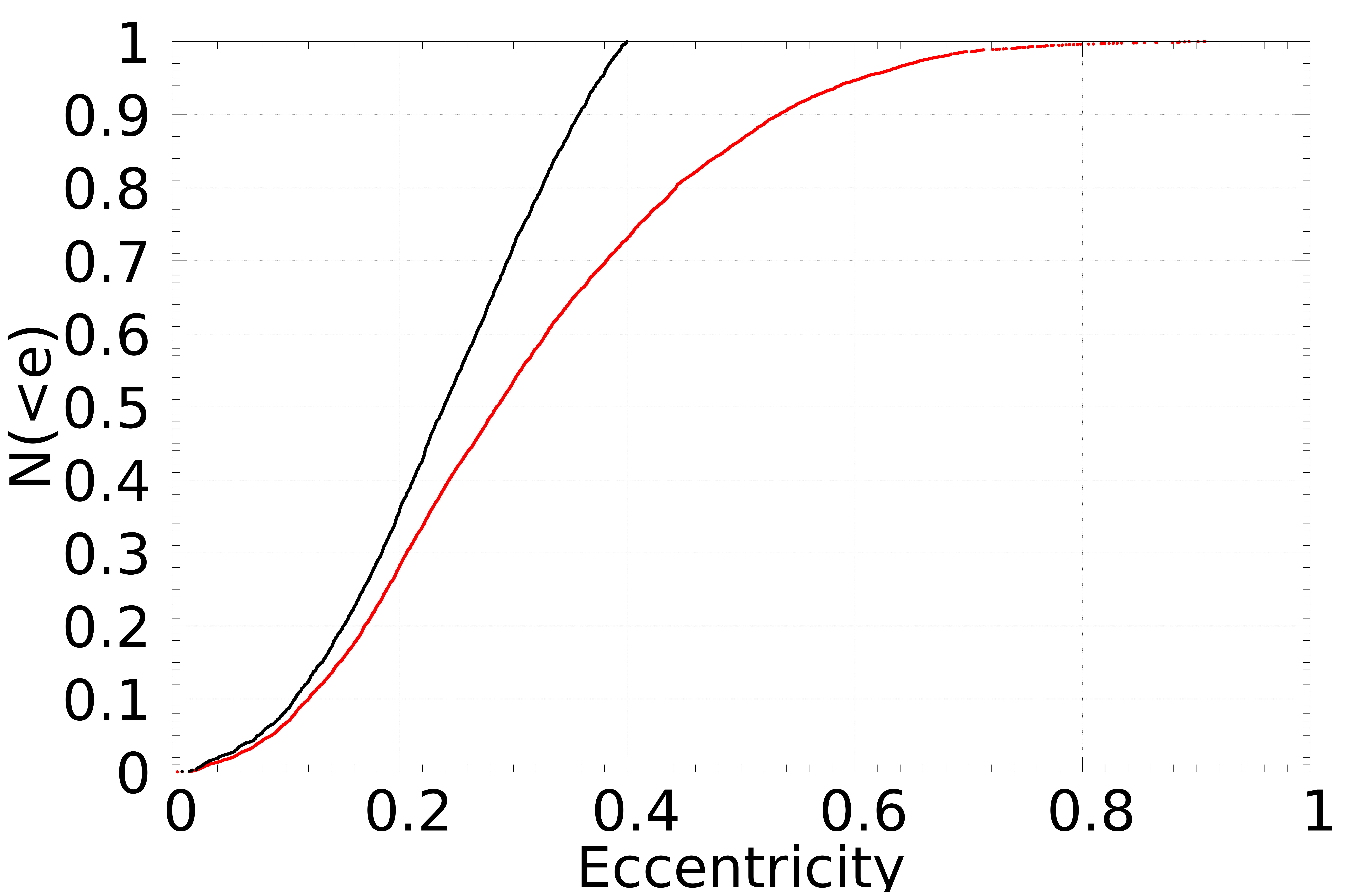} {0.3\textwidth}{(b)}
\fig{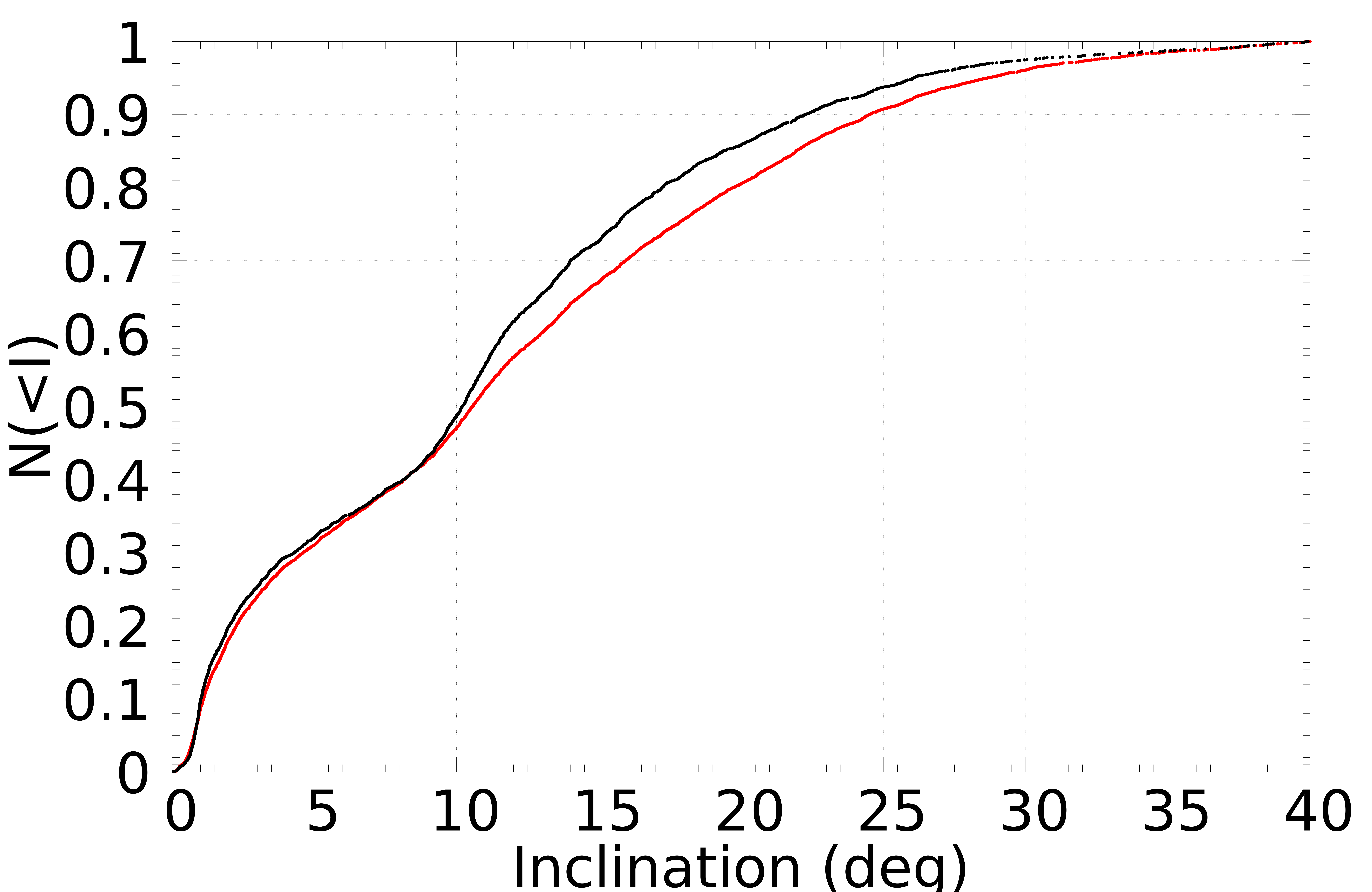} {0.3\textwidth}{(c)} }

\caption{Panels (a), (b) and (c) show the cumulative distribution of semimajor axis, eccentricity and inclination of the implanted planetesimals. 
The red color and black color stand for the cumulative distribution at the end of the gas phase for the inner Solar System region (a $<$ 3 au) and for the region with semimajor axis between 2 and 3.0 au, eccentricity 
smaller than 0.4 and inclination smaller than 40$^{\circ}$, respectively.}  
\label{fig:black122}
\end{figure*}

The cumulative normalized distribution of the semimajor axes, eccentricities and inclinations of the cloned planetesimals at the end of our simulations are shown in
Panels (a), (b) and (c) by red curves in Figure \ref{fig:black122}. 
Note that 40\% of the cloned planetesimals that survived untill the end of the simulation, have a semimajor axis between 2.5 and 3 au, 25\% have an eccentricity smaller than 0.2 and 
about 45\% have an orbital inclination smaller than 10$^{\circ}$. 
If we consider only planetesimals captured in the region with 2.0 $<$ a $<$ 3 au, e $<$ 0.4 and I $<$ 40$^{\circ}$, the cumulative normalized distribution of a, 
e and I are shown in Panels (a), (b) and (c) by the black curves. In this case, note that 70\% of the planetesimals captured in this region 
have semimajor axis between 2.5 and 3 au, 33\% have an eccentricity smaller than 0.2 and 45\% have an inclination 
orbit smaller than 10$^{\circ}$. 

\subsection{Capture Results in the asteroid belt region}
\label{capturer1}

We evaluate which fraction of Ceres candidates are permanently stored in the asteroid belt in our simulations. 
Similarly to \citet{2016AJV}, we use the mass of the disk and size distribution proposed in \citet{Nesvorny2016a} to estimate the 
probability of a Ceres permanent capture at the end of the gas disk phase. 

The capture efficiency ($C_E$), or probability, per planetesimal is calculated using: 

\begin{equation} 
C_E= r_{region} \times C_{EM},
\end{equation}

\noindent where $r_{region}$ is the ratio between the number of planetesimals implanted in stable orbits in a particular region of the asteroid belt and 
the total number of cloned planetesimals while $C_{EM}$ is the fraction of disk planetesimals that became \textit{Ceres candidate} and have been cloned.

Of the total number of 294,087 clones, we found 3012 bodies captured in stable orbits in the asteroid belt between 2.1 and 3.25 au 
and eccentricity smaller than 0.4. Thus, the capture efficiency in the asteroid belt per \textit{Ceres candidate} is $\frac{3012}{294,087}$.
We assume that this probability is the same whatever the original provenance in the trans-Saturnian disk. Because the probability that a planetesimal starting 
from the black portion of the disk becomes a \textit{Ceres candidate} is $1.11 \times 10^{-1}$, the probability of implantation in the asteroid belt from this portion of the disk is $C_{E}= \frac{3012}{294,087} \times 1.11 \times 10^{-1} = 1.20 \times 10^{-3}$. Repeating the calculation for planetesimals in the red and blue portions of the disk (see Table \ref{tab:mathmode0} 
for the corresponding values of $C_{EM}$) we find $4.36 \times 10^{-4}$ and $2.87 \times 10^{-5}$ respectively.  
The probability of capture in the inner asteroid belt (a $<$ 2.5 au) is 8/30 of the total. 

Assuming that there were  3600--14,000 Ceres-sized bodies in each of the three considered portions of the protoplanetary disk, and 
using the capture efficiencies for each region, the probabilities to implant a Ceres-like planetesimals in the asteroid belt are: 4-17, 1-6, 0.10-0.40 from the black, red and blue portions of the disk respectively.
A probability larger than 1 means that more than one captured object is expected.
How these probabilities are partitioned between the inner, middle and outer parts of the belt is shown in Table \ref{tab:mathmode4}. 

\section{Discussion and Conclusions}
 \label{discussion}
 
The results of our numerical simulations support the idea that some Ceres-sized planetesimals might have been implanted from the trans-Saturnian region into the asteroid belt by
a combination of close encounters with giant planets, with other protoplanets, gas drag, chaotic diffusion and mean motion resonances with Jupiter. In this section we discuss how our result fits within a broader context. 
We describe how the scenario proposed here can fit a narrative proposed for the asteroid belt evolution: during the gas phase (Section \ref{astevol}), after the gas phase (Section \ref{astevolafter}) and the total mass implanted in 
the asteroid belt (Section \ref{astevolmass}). We conclude with a discussion of the limitations of our work in Section \ref{limit}.

\subsection{Asteroid belt evolution during the gas phase}
\label{astevol}

Ceres and the other planetesimals were probably formed during the first five million years after the formation of the first solids in the inner Solar System (CAIs \citep{2010NatGeBouvier}). 
The quick formation of planetesimals can be explained by the radial pile-up of dust, reaching a sufficient concentration to trigger the streaming instability \citep{2005YoudinGoodman,2017JohasenLambrechts,2018MNRASquireHopkin}. 
The streaming instability is expected to form  planetesimals with most of the mass concentrated in large objects (hundreds of km in size) and, in addition, the largest planetesimals can continue to grow by accreting other 
planetesimals \citep{2000kokuboida} or radially drifting dust \citep{lambrechts12}.

The first phase of the Asteroid Belt evolution happens during the gaseous disk phase. Jupiter and Saturn formed close to the asteroid belt. While the S-type asteroids probably formed in-situ, 
the C-type planetesimals were implanted from the Jupiter-Saturn region by the combined effects of giant planet growth, orbital destabilization, gravitational scattering and aerodynamic gas drag
\citep{raymond17,2020RaymondNesvorny}. 

While considering the Raymond et al. scenario for the origin of C-type asteroids, we propose another possible scenario for the implantation of Ceres in the asteroid belt. 
We remind the reader that Ceres is not C-type and is rich in ammonia compounds that are typically not found in C-type asteroids and carbonaceous chondrites. This motivates us to look for 
an alternative mechanism capable of implanting Ceres from farther out in the disk than the original source region of C-type asteroids.

In our model, the implantation of a Ceres-size planetesimal started at the time when Jupiter and Saturn became giant planets and a system of protoplanets of about
5 $M_\oplus$ formed beyond Saturn orbit. The migration of the protoplanets blocked by Jupiter and Saturn produced by mutual collisions a set of ice giant planets with Neptune-like masses ($m \sim 15 M_\oplus$) \citep{Izidoroetal2015}.  
In this work, we show that 11\% of the total population of Ceres-size objects initially between 10 and 20 au is temporarily trapped in the asteroid belt as a \textit{Ceres candidate}, due to the combined scattering from the protoplanets and the giant planets. 
A smaller fraction (0.28\%) is implanted  from 30 to 50 au region. This fraction is not zero, thanks to the partial radial mixing of planetesimals in the trans-Saturnian disk due to close encounters with the migrating protoplanets. 

In order to improve statistics and calculate the probability that a Ceres-like planetesimal is permanently implanted into the asteroid belt region, we clone 500 times all the \textit{Ceres candidates}. 
We show that the temporary population of Ceres-size planetesimals can reach stable orbits by three different mechanisms: capture into mean motion resonances with Jupiter, 
scattering off protoplanets temporarily crossing the inner solar system and gas drag. Some permanently implanted Ceres-size planetesimals can be found close to the current orbit of Ceres.

We find that a fraction of about 1\% (3012 planetesimals) of the Ceres candidates survives and reaches the asteroid belt 
region defined by $2.1 < a < 3.25$ au and eccentricity smaller than 0.4 (Table \ref{tab:mathmode}). 
We also determine the capture efficiencies separately for each region of the asteroid belt and for each natal region in the trans-Saturnian planetesimal disk (black, red, and blue regions; Table \ref{tab:mathmode3}). 
The capture efficiency in the asteroid belt region per disk of planetesimals varies from $2.85 \times 10 ^{-5}$ to  $1.20 \times 10 ^{-3}$ for the outer and inner parts of the trans-Saturnian disk, respectively. 

We use the mass of the planetesimal disk and its size distribution proposed in \citet{Nesvorny2016a} to estimate the probability that Ceres was captured by this process. We assume that the primordial disk beyond Saturn 
contained $3600-14000$ Ceres-sized planetesimals \citep{Nesvorny2016a}. Combining these numbers with our implantation probabilities, (see Table \ref{tab:mathmode4}), we find that about 4-17 Ceres-size objects could have been implanted in the asteroid belt if they started in the inner region of the disk. 
Instead, the probability to implant a Ceres-sized objects from the outer region of the disk (a $>$30 au) drops to 10-40\% (in addition it is unlikely that the disk beyond 30 au contained thousands of Ceres-size objects given that the cold Kuiper belt contains none).
 Therefore, we argue that Ceres is more likely to come from the inner region of the trans-Saturnian disk, where the number of Ceres-sized objects implanted in the asteroid belt can be greater than 1. If we look at our normalized and cumulative orbital distribution of the implanted planetesimals, 
we find that 33\% of the implanted planetesimals have eccentricity smaller than 0.2 and about 45\% of them have a probability of having an inclination smaller than 10 degrees. We also found that 70\% of the implanted planetesimals have a semimajor axis between 2.5 and 3 au. 

Looking at how many Ceres-sized planetesimals are implanted in other regions that do not have a Ceres-like object today, for instance the inner and outer asteroid belt region, we find 1-6 and 2-10 Ceres-like objects, 
respectively from red and black regions. Therefore, in our scenario, it is more probable to have Ceres-like planetesimals in the outer asteroid belt region than in the central asteroid belt region. We discuss a possible solution of 
this discrepancy in Sect.~\ref{astevolafter}. 

\subsection{Asteroid belt evolution after the gas phase}
\label{astevolafter}

Now we turn our discussion to the asteroid belt's dynamical evolution after the disappearance of the gas disk. During this period, some planetesimals may be implanted from the terrestrial planet 
region scattered by the population of Mars-mass protoplanets which formed the terrestrial planets \citep{2006IcarRaymondSean,Morbietal2012,2017SciARaymondIzidoro}. However, some planetesimals 
may have been implanted from the trans-Neptunian region during the giant planet instability \citep{2009NaturL,2016AJV}. The planetesimals implanted from the trans-Neptunian region during the
instability phase are usually identified with P/D type asteroids, because these are the types that dominate in the Jupiter Trojan population, 
which is expected to have been implanted at the same time from the same source region \citep{Nesvornyetal2013}.  

The probability of capturing Ceres from the trans-Neptunian region during the giant planet instability was calculated by \citet{2016AJV}. They also considered that the primordial disk beyond Neptune 
between 23 and 30 au may have had 3600 - 14000 Ceres-sized planetesimals. However, their computations showed that the capture probability in this phase is only $3\times10^{-7}$ per disk planetesimal, so 
that the probability to capture a Ceres-like object is only is 0.1 to 0.4\%. They concluded that this probability is too low to support the capture of Ceres from the trans-Neptunian region during the giant planet instability phase.
This makes sense, because capture typically requires sufficient gas drag after the transport, which is nonexistent once the primordial gas dissipates.

The giant planet instability can also partially deplete the asteroid belt, including the pre-implanted objects. The depletion of the asteroid belt depends on the asteroid belt initial orbital distribution and on the 
specific evolution of the planets. \citet{Roigenesvorny2015} and \citet{Nesvornyetal2017} studied the asteroid belt evolution during the giant planet instability. \citet{Nesvornyetal2017} found that about 20\% of the 
original asteroids (starting at 2 to 3.5 au) would survive in the main belt after the giant planet instability (i.e. and 80\% depletion). \citet{Clement2019b} tested the possibility that the giant planet  orbital instability depleted more substantially 
the main belt, leading to its small current mass and orbital structure. In some cases \citet{Clement2019b} found depletion factors for the main belt larger than 99\%. Therefore, the depletion factor of the asteroid belt during the instability is still very uncertain.

The number of Ceres-sized planetesimals implanted in our scenario (Table \ref{tab:mathmode4}) is consistent with depletion factors during the giant planet instability of 75-94\%.
If larger than 94\%, our model does not expect that an implanted Ceres-size planetesimal survives in the asteroid belt today. If less than 75\%, it predicts too many Ceres-sized planetesimals surviving in the asteroid belt.
Thus, our scenario is consistent with the depletion factors found in \citet{Roigenesvorny2015} but it is not consistent with the extreme depletion factor found in \citet{Clement2019b}.

Notice that the dynamical depletion of the asteroid belt during the giant planet instability is typically stronger in the outer belt than in the central belt. Thus, our result that the trapping probability in the 
outer belt is the highest may not be in contradiction with the observation that the only
surviving Ceres-like body is in the central belt. 

\subsection{Total mass implanted in the Asteroid Belt}
\label{astevolmass}

Because there is gas in our simulations, the implantation efficiency is probably higher for smaller planetesimals, due to the effect of gas drag. However, assuming for simplicity that that the implantation efficiency is size independent and assuming that the trans-Saturnian disk
comprised at least 30 Earth masses (this is the mass needed in the trans-Neptunian disk in the \textit{Nice} model), we estimate that 3--12$\times10^{-3}$ Earth masses of trans-Saturnian planetesimals should have been implanted into the asteroid belt. This is
much more than the total mass of the asteroid belt.

The problem arises from the fact that Ceres comprises $1/3$ of the total mass of the asteroid belt, which is much larger than the fraction of the disk mass in Ceres-like planetesimals according to the size-frequency distribution (SFD) considered by \citet{Nesvorny2016a}
and observed today in the hot Kuiper belt population. Thus, a model that implants in the asteroid belt the correct number of Ceres-like planetesimls necessarily implants a too much in total mass, whereas a model that implants an acceptable total mass necessarily implants an insufficient number of Ceres-like planetesimals. 

The simplest possible solution to this conundrum seems to be that Ceres was derived from a region of the disk that had a top-heavy SFD, with a Ceres-mass/total-mass ratio more similar to that of the asteroid belt than to that of the Kuiper belt.
We speculate that this SFD characterized the planetesimals in the vicinity of Saturn, whereas the SFD of \citet{Nesvorny2016a} characterized the more distant disk which, as we have seen, has a very small probability to implant objects in the asteroid belt.  

\subsection{Limitations of our work}
\label{limit}

As with any numerical study, our simulations are a simplified and idealized version of reality. One main limitation is that our results are based on a single simulation from \citet{Izidoroetal2015}. 
Nevertheless, this simulation represents, to our knowledge, the most successful one in reproducing the origin of Uranus and Neptune as well as of a third ice-giant planet, required in the most 
recent versions of the \textit{Nice} model. 
However, the implantation of objects into the Asteroid Belt also depends on the scattering process caused by the protoplanets. The scattering process, therefore, depends on how long the protoplanets stay 
in the disk, the number of them (in particular of the invaders) and the migration of the whole chain.
Thus, other formation histories of the ice giant planets, even within the broad \citet{Izidoroetal2015} model, could possibly lead to different results 
in terms of efficiency of implantation of objects in the main belt. In particular, we did not consider a case where no protoplanet is temporarily 
scattered onto an orbit interior to Jupiter’s (\textit{invader}), because the production of invaders is a characteristic feature of the \citet{Izidoroetal2015} model. 
Another factor that could be important is the density of the gas that could change the role of gas drag and hence the probabilities to capture objects 
in the inner Solar System. To be consistent with the planetary evolution of \citet{Izidoroetal2015}, we used only one profile for the gas density. 
The total number of Ceres-like bodies we assumed in the trans-Saturnian disk is inspired by the work of \citet{Nesvorny2016a}, which is valid only for the trans-Neptunian region at the end of the planet formation phase, 
so it also suffers from large uncertainties. Finally, we did not consider the terrestrial planets formation in our simulations, which is reasonable, in first approximation. However, there might have been protoplanets 
forming in the inner Solar System during the period of time we considered. These local protoplanets might also have affected the implantation probability and the survival in the Asteroid belt region.
Another caveat in our study is the fact that we have only one real Ceres in the asteroid belt, therefore it is subject to small number statistics. Ceres could have followed a 
special track in the dynamical evolution of the Solar System. We are considering here that if a model gives less than 4.6\% of probability (which in statistics is like 2.0 $\sigma$ for a
normal probability distribution), we must look for another model that explains the existence of the object with reasonable probability.

\section*{Acknowledgments} 
We thank the anonymous referee for the valuable comments andsuggestions that helped improve the quality of this paper. 
This work was possible thanks to the scholarship granted from the Brazilian Federal Agency for Support and Evaluation of Graduate Education (CAPES), in the scope of the Program CAPES-PrInt, process number 88887.310463/2018-00, 
Mobility number {88887.572647/2020-00} {88887.468205/2019-00}.
This  research  was  supported in  part  by the S\~ao Paulo Research Foundation (FAPESP) through the  computational resources provided  by  the  Center for Scientific Computing (NCC/GridUNESP) of the S\~ao Paulo State University (UNESP).
Ribeiro  acknowledges  support  provided  by  grants No. 2016/24561-0, S\~ao Paulo Research Foundation (FAPESP).
A. Izidoro  acknowledges NASA grant 80NSSC18K0828 (to Rajdeep Dasgputa) for financial support during preparation and submission of the work.
A. Izidoro acknowledges support from the Welch Foundation grant No. C-2035-20200401.
A. Izidoro thanks also support from CNPq via process 313998/2018-3 and CAPES/PrInt/UNESP  during initial development of this project.


\bibliography{bib}{}
\bibliographystyle{aasjournal}



\end{document}